\documentclass{article}
\usepackage[utf8]{inputenc} %Allows accenting and some direct symbol typesetting
\usepackage{xcolor} %Allows us to use text colors
\usepackage{soul} %Allows us to highlight text
\usepackage{fancyhdr} %Enables fancy headers
\usepackage{xcolor} %Enables colored text.

\usepackage{comment} %Allows comment environment.
\usepackage{authblk} %Creates author blocks - more flexible author affiliations.
\usepackage{blindtext} %Creates well-sampled 
\usepackage[tmargin=1.0in,bmargin=1.0in,lmargin=1in,rmargin=1in]{geometry} %Set page geometry.
\usepackage{graphicx} %Allows figure input
\usepackage{float} %Improves control over floating objects
\usepackage[backend=biber,style=phys]{biblatex} %Enables bibliography formatting and styling.
\usepackage{hyperref} %Allows hyper-referencing.
\hypersetup{
    colorlinks=true, %set true if you want colored links
    linktoc=all,     %set to all if you want both sections and subsections linked
    linkcolor=blue,  %choose some color if you want links to stand out
    citecolor=blue,
    urlcolor=blue
}
\usepackage{mathtools}
\usepackage{matlab-prettifier}
\usepackage{multicol}
\usepackage{wrapfig}
\usepackage[export]{adjustbox}
\usepackage{graphicx} % Required for inserting images
\title{Electric Fields Near Undulating Dielectric Membranes}
\author[1]{Nicholas Pogharian}
\author[1,2]{Alexandre P. dos Santos}
\author[3]{Ali Ehlen}
\author[1,3,4,5,6]{Monica Olvera de la Cruz \footnote{Corresponding author: m-olvera@northwestern.edu}}
\affil[1]{Department of Materials Science and Engineering, Northwestern University, Evanston, Illinois, 60208, USA}
\affil[2]{Instituto de F\'{i}sica, Universidade Federal do Rio Grande do Sul, 91501-970 Porto Alegre, RS, Brazil}
\affil[3]{Applied Physics Program, Northwestern University, Evanston, Illinois 60208, USA}
\affil[4]{Department of Physics and Astronomy, Northwestern University, Evanston, Illinois 60208, USA}
\affil[5]{Department of Chemistry, Northwestern University, Evanston, Illinois 60208, USA}
\affil[6]{Department of Chemical and Biological Engineering, Northwestern University, Evanston, Illinois 60208, USA}

\begin{document}

\maketitle
\begin{abstract}
    Dielectric interfaces are crucial to the behavior of charged membranes, from graphene to synthetic and biological lipid bilayers. Understanding electrolyte behavior near these interfaces remains a challenge, especially in the case of rough dielectric surfaces. A lack of analytical solutions consigns this problem to numerical treatments. We report an analytic method for determining electrostatic potentials near curved dielectric membranes in a two-dimensional periodic 'slab' geometry using a periodic summation of Green's functions. This method is amenable to simulating arbitrary groups of charges near surfaces with two-dimensional deformations. We concentrate on one-dimensional undulations. We show that increasing membrane undulation increases the asymmetry of interfacial charge distributions due to preferential ionic repulsion from troughs. In the limit of thick membranes we recover results mimicking those for electrolytes near a single interface. Our work demonstrates that rough surfaces generate charge patterns in electrolytes of charged molecules or mixed-valence ions.
\end{abstract}
\section{Introduction}
Dielectric interfaces are ubiquitous in the physical world. They are essential to a wide variety of important systems including biological proteins\supercites{bibi_review_2016}{li_dielectric_2013}{schutz_what_2001}{loffler_calculation_1997}{warshel_modeling_2006}{kukic_protein_2013}, electrochemical capacitors \supercites{varner_effects_2022}{molina-reyes_design_2018}{platonov_study_2021}{floch_fundamental_2020}, quantum dots\supercite{franceschetti_addition_2000}, clays\supercites{zhang_frequency_2020}{hubbard_estimation_1997}, and ionomers\supercite{kwon_determining_2019}. These interfaces are especially important in nanoconfined\supercites{kesselheim_effects_2012}{bagchi_surface_2020}{zwanikken_tunable_2013}{jimenez-angeles_surface_2023}{nguyen_manipulationof_2019} systems, influencing the development of new electronic materials in the field of iontronics. Due to the dielectric mismatch at these interfaces, it is challenging to understand exact electrostatic behavior in their vicinity. One typical way to study this behavior is the Poisson-Boltzmann equation, which renders the mathematics more analytically tractable by removing correlations between charged particles and ignoring finite-size effects. While exact solutions exist for simple, flat geometries, only approximate solutions exist for more complex interfaces.\supercites{behrens_exact_1999}{zhang_exact_2018}{xing_poisson-boltzmann_2011} Even for exact solutions, errors due to assumptions of the Poisson-Boltzmann theory become relevant at high charge concentrations where charge correlations become important.\supercites{levin_electrostatic_2002}{netz_electrostatistics_2001}{adar_screening_2019} There have been attempts made to extend Poisson-Boltzmann theory to include correlations and finite-size effects using methods such as hypernetted chain theory and charge renormalization, but these remain accurate only in certain regimes such as those of low fluctuations.\supercites{borukhov_steric_1997}{lozada-cassou_application_1982}{henderson_recent_1983}{ding_charged_2016}{buyukdagli_beyond_2016} In addition to exact Poisson-Boltzmann theory, methods employing a linearized Poisson-Boltzmann (Debye-H\"{u}ckel) theory have been able to successfully describe the electric double layer near bent interfaces. These efforts led to both perturbative and iterative approximate solutions to the linearized Poisson-Boltzmann equation describing deformed membranes, as well as exact solutions for closed surfaces using multiple scattering expansions.\supercites{goldstein_electric_1990}{duplantier_geometrical_1990}. This work additionally allowed for determinations of electric double layer free energy for rough fractal surfaces.\supercite{goldstein_thermodynamics_1991}\par
In addition to theoretical investigation, experimental characterization of electric double layers arising from dielectric mismatch has been attempted; this also proves challenging.\supercites{favaro_unravelling_2016}{brown_measure_2013} Due to these experimental and theoretical difficulties, numerical methods are often the only way to study these systems. Typically, these methods rely on computing the charge accumulated at dielectric interfaces based on the discontinuity of the normal gradient of the electrostatic potential.\supercites{limmer_charge_2013}{merlet_electric_2014}{jing_ionic_2015}{boda_computing_2004}{gan_comparison_2015}{tyagi_iterative_2010}{gan_efficient_2019}{yuan_particleparticle_2021} While they can be effective, they are often quite slow, limiting simulations to short timescales and small numbers of particles. To circumvent this limitation, dos Santos and Levin have in the past developed methods considering periodic image charge contributions in two-dimensional periodic 'slab' geometries.\supercite{santos_electrolytes_2015} Building upon these ideas, they have also developed methods which take advantage of periodic Green's functions to solve the Poisson equation and calculate polarization contributions to the electrostatic potential for systems in these slab geometries.\supercites{girotto_simulations_2017}{santos_simulations_2017}{santos_dielectric_2018} These calculations are typically faster than numerical methods while additionally being analytical. Although these methods require a summation over all periodic images of the system, they converge very quickly in practice.\par
In this work, we introduce a first-order perturbation scheme to calculate electrostatic potentials arising from a system of point charges near a bent membrane interface in a system with periodicity in two dimensions. Using this method, we present theoretical self-interaction energies of particles near a dielectric interface. In addition to these single-particle results, we consider systems of highly concentrated electrolytes in the vicinity of a rough interface using Langevin dynamics. Highly concentrated electrolytes in confinement are particularly important in neuromorphic applications as well as in the field of iontronics.\supercites{robin_modeling_2021}{kamsma_iontronic_2023}{han_iontronics_2022}{khan_advancement_2023} We study systems with divalent positive and monovalent negative ions to explore charge asymmetry effects, additionally varying the interfacial roughness to determine the impact of surface curvature on electrolyte structure.
\section{Model and Calculations}
\label{section:model}
Our system consists of charges $Q$ in some dielectric medium near a membrane of thickness $2d$. The dielectric constant outside of the membrane is $\epsilon_1$, while the dielectric constant inside the membrane is $\epsilon_2$. This system is considered to be infinitely periodic in the x and y directions, an example of so-called 'slab' geometry. The volume of this periodic box is $L_xL_yL_z$. This membrane problem is readily amenable to solutions using previously developed methods using periodic Green's functions.\supercite{santos_simulations_2017} The problem at hand is complicated by the fact that the membrane surface is not perfectly flat; it is subject to some small deformation $s(x,y)$. Since this deformation is periodic, we cast it in the form $s(x,y)=He^{i(q_xx+q_yy)}$ as shown in Equation \ref{eqn:h definition} where $H$ is the amplitude of the deformation and $q_x$ and $q_y$ are the inverse wavelengths of undulation. 
\begin{figure}[H]
    \centering
    \includegraphics[width=0.5\columnwidth]{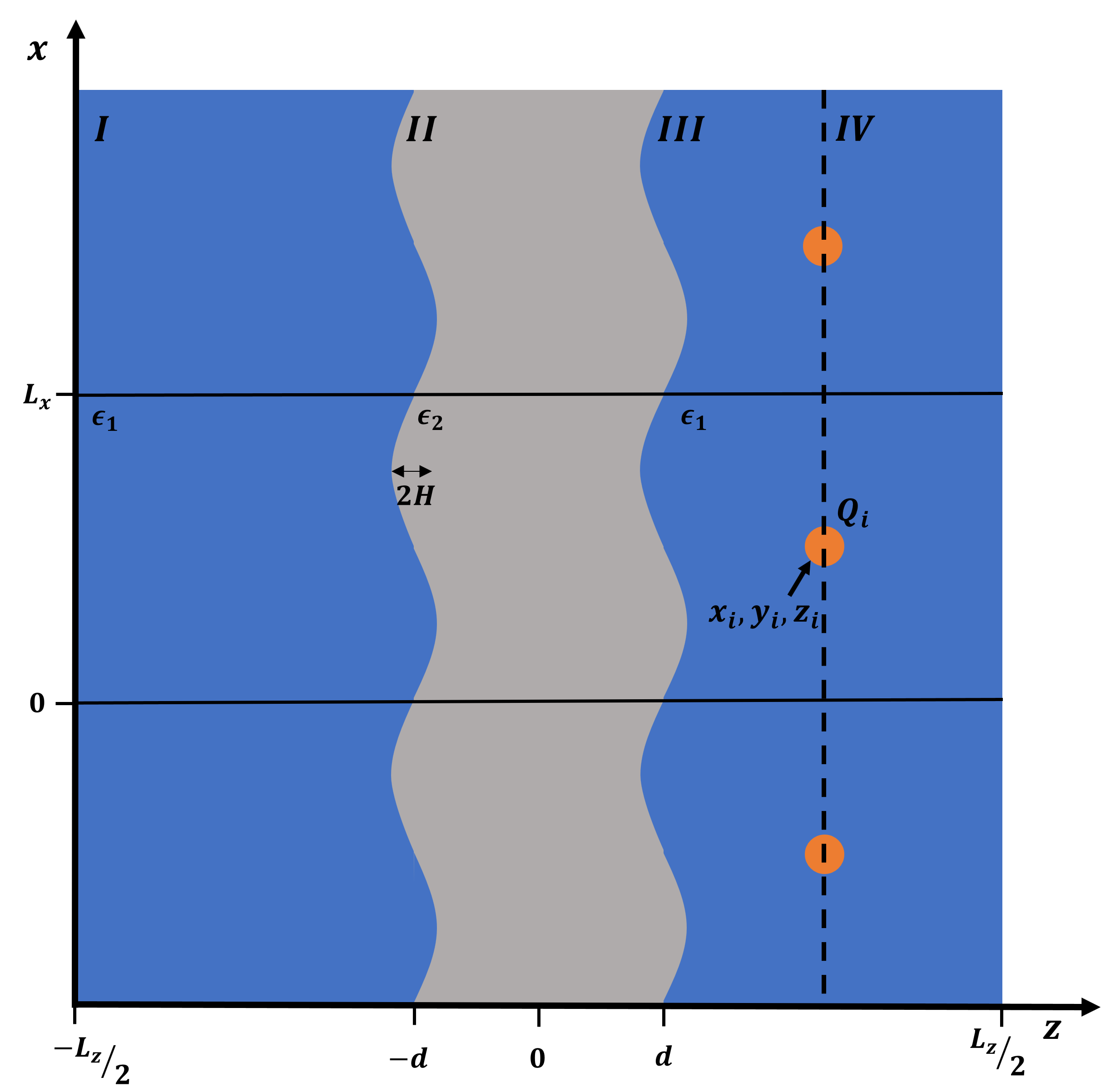}
    \caption{Schematic of the system. An ion with charge $Q_i$ is in the vicinity of a flat membrane subjected to some small deformation $s(z)$ of amplitude $H$. The dielectric constant inside the membrane is different than that outside. The horizontal black lines show the periodicity of both the system and the perturbation to the flat surface. The system is additionally divided into four regions.}
    \label{fig:box with regions}
\end{figure}
To describe the electrostatic potential throughout the system, we must solve the Poisson equation:
\begin{equation}
\label{eqn:poisson}
    \nabla^2 \phi = -\frac{4 \pi \rho}{\epsilon_1},
\end{equation}
where $\phi$ is the electrostatic potential, $\rho$ is the charge density, and $\epsilon_1$ is the dielectric constant of the system. Rather than immediately considering the collective potential of all charges, it is simpler to consider the potential from a single point charge and its periodic images in $x$ and $y$ as shown in Figure \ref{fig:box with regions}. We consider this point charge to have charge $Q_i$ and position $\vec{r_i}=(x_i,y_i,z_i)$. This results in the following expression:
\begin{equation}
\label{eqn:poisson with green function}
    \nabla^2 G(\vec{r},\vec{r_i}) = -\frac{4 \pi Q_i}{\epsilon_1} \sum_{m_x,m_y=-\infty}^{\infty} \delta (\vec{r}-\vec{r_i} + m_x L_x \hat{x} + m_y L_y \hat{y}),
\end{equation}
where $G(\vec{r},\vec{r_i})$ represents the Green's function for the system. The periodic delta functions in x and y can be expressed in the Fourier transform representation such that we obtain the following equation:
\begin{equation}
\label{eqn:green fn FT representation}
    \nabla^2 G(\vec{r},\vec{r_i}) = -\frac{4 \pi Q_i}{\epsilon_1L_xL_y} \delta (z-z_i) \sum_{\vec{m}=-\infty}^{\infty} e^{2 \pi i \left[\frac{m_x}{L_x}(x-x_i)+\frac{m_y}{L_y}(y-y_i)\right]},
\end{equation}
where $\vec{m}=(m_x,m_y)$.
To solve Equation \ref{eqn:green fn FT representation}, we employ the following \textit{ansatz}:

\begin{equation}
    \label{eqn:ansatz}
    G(\vec{r},\vec{r_i})=\frac{1}{L_xL_y} \sum_{\vec{m}=-\infty}^{\infty} \left(g_{\vec{m}}^{(0)}(z,z_i) + h(x,x_i,y,y_i)g_{\vec{m}}^{(1)}(z,z_i)\right) e^{2 \pi i \left[\frac{m_x}{L_x}(x-x_i)+\frac{m_y}{L_y}(y-y_i)\right]}
\end{equation}

\begin{figure}[H]
    \centering
    \includegraphics[width=\columnwidth]{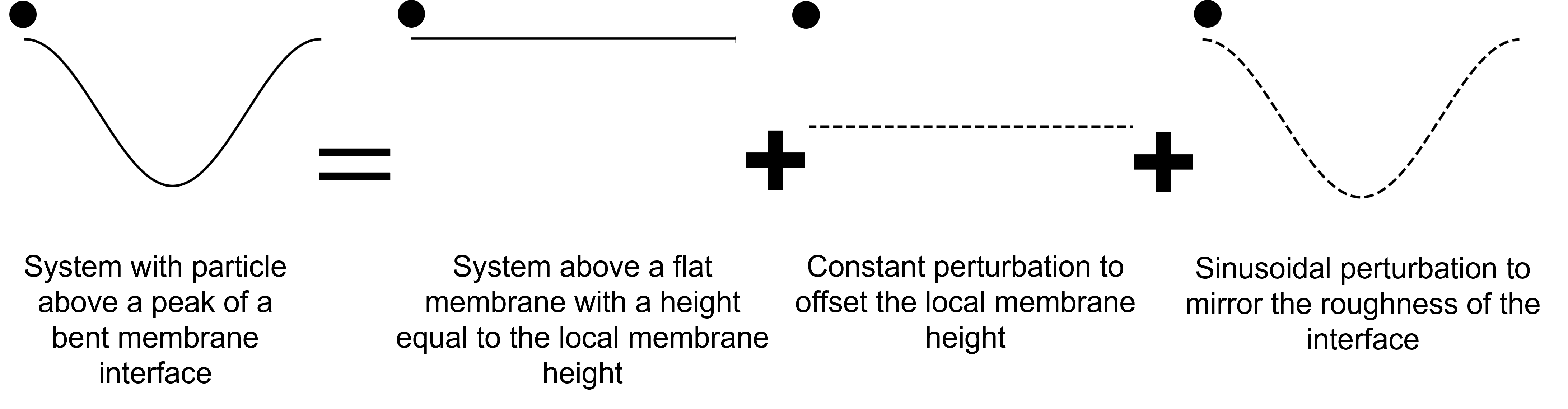}
    \caption{Schematic of the perturbative scheme. A particle above a peak in a rough interface is imagined as being above some flat surface with a location corresponding to the local height of the particle above the membrane. This is modified by both constant ($h_c$) and sinusoidal ($h_s$) perturbations to approximate particle behavior near the peak of a deformed interface.}
    \label{fig:perturbative scheme schematic}
\end{figure}

We take a perturbative approach of expanding the Green's function to first order in the small parameter $h$, which we can use to reduce the problem to a set of coupled differential equations and boundary conditions as shown in perturbative solutions to the Debye-H\"{u}ckel equation.\supercite{goldstein_electric_1990} The zero-order contribution to the Green's function is denoted as $g_{\vec{m}}^{(0)}$, while the first order contribution is represented by $hg_{\vec{m}}^{(0)}$. Since we are considering a sinusoidal deformation of the surface, it appears natural to express our potential as a sinusoidal perturbation $h(x,y)=He^{i(q_xx+q_yy)}$ about a flat membrane. This is a useful approximation for particles far from the membrane; however this does not accurately describe particles close to the membrane surface.\supercite{solis_pimples_2021} For this reason, instead of expanding around $z=\pm d$, we expand around $z=\pm d+h_0$ where $h_0$ is the local undulation height at the particle position. For a sinusoidal deformation of amplitude $H$ taking the form $s(x,y)=He^{i(q_xx+q_yy)}$, the potential felt from a particle above a peak would be expanded around $z=d+H$ using a perturbation $h(x,y)=H(e^{i(q_xx+q_yy)}-1)$. For a trough, this would take the form of an expansion around $h_0=-H$ using a perturbation $h(x,y)=H(e^{i(q_xx+q_yy)}+1)$. This concept is illustrated in Figure \ref{fig:perturbative scheme schematic}.
Based on this concept of splitting the perturbation into sinusoidal and constant components, we define the perturbation $h$ in the following way:
\begin{subequations}
\label{eqn:h definition}
\begin{equation}
    h=H\left(e^{i(q_xx+q_yy)}-e^{i(q_xx_i+q_yy_i)}\right)=h_s+h_c,
\end{equation}
\begin{equation}
    h_s=He^{i(q_xx+q_yy)},
\end{equation}
and
\begin{equation}
    h_c=-He^{i(q_xx_i+q_yy_i)},
\end{equation}
\end{subequations}
where $H$ is the amplitude of the deformation, $x_i$ and $y_i$ represent the $x$ and $y$ positions of particle $i$, and $x$ and $y$ represent the coordinates where we wish to evaluate the electrostatic potential $\phi$ The perturbation $h$ can be further divided into two separate terms $h_s$ and $h_c$; these represent the sinusoidal and constant components of the deformation respectively. Plugging our expressions into Equation \ref{eqn:green fn FT representation}, we obtain the following:
\begin{equation}
\begin{multlined}
    \label{eqn:ansatz plus substitution full}
    \frac{1}{L_xL_y}\sum_{\vec{m}=-\infty}^{\infty} \left[\frac{\partial^2 g_{\vec{m}}^{(0)}(z,z_i)}{\partial z^2}-k^2g_{\vec{m}}^{(0)}(z,z_i) + h_s\left(\frac{\partial g_{\vec{m}}^{(1,s)}(z,z_i)}{\partial z^2} -\kappa^2g_{\vec{m}}^{(1,s)}(z,z_i)\right) \right. \\ \left.+ h_c\left(\frac{\partial g_{\vec{m}}^{(1,c)}(z,z_i)}{\partial z^2} -k^2g_{\vec{m}}^{(1,c)}(z,z_i)\right)\right]e^{2 \pi i \left[\frac{m_x}{L_x}(x-x_i)+\frac{m_y}{L_y}(y-y_i)\right]}= \\ -\frac{4 \pi Q_i}{\epsilon_1L_xL_y} \delta (z-z_i) \sum_{\vec{m}=-\infty}^{\infty} e^{2 \pi i \left[\frac{m_x}{L_x}(x-x_i)+\frac{m_y}{L_y}(y-y_i)\right]},
\end{multlined}
\end{equation}
where $k=2\pi \sqrt{m_x^2 / L_x^2 + m_y^2 / L_y^2}$ and $\kappa=\sqrt{(q_x+2 \pi m_x / L_x)^2+(q_y+2 \pi m_y / L_y)^2}$. The first-order contribution to the Green's function $hg_{\vec{m}}^{(1)}$ is further decomposed into $h_sg_{\vec{m}}^{(1,s)}$ and $h_cg_{\vec{m}}^{(1,c)}$, representing the constant and sinusoidal components of the deformation respectively. We solve this equation by considering each term in the summations over $\vec{m}$ separately. Moreover, we consider independently the terms proportional to $h^0$, $h_s^1$, and $h_c^1$. This leads to the following three equations which can be evaluated to determine the electrostatic potential:
\begin{subequations}
\begin{equation}
    \label{eqn:zeroth order equation}
    \frac{\partial^2 g_{\vec{m}}^{(0)}(z,z_i)}{\partial z^2}-k^2g_{\vec{m}}^{(0)}(z,z_i)=-\frac{4 \pi Q_i}{\epsilon_1L_xL_y} \delta (z-z_i),
\end{equation}
\begin{equation}
    \label{eqn:first order equation}
    \frac{\partial^2 g_{\vec{m}}^{(1,s)}(z,z_i)}{\partial z^2}-\kappa^2g_{\vec{m}}^{(1,s)}(z,z_i)=0,
\end{equation}and
\begin{equation}
    \label{eqn:first order equation c term}
    \frac{\partial^2 g_{\vec{m}}^{(1,c)}(z,z_i)}{\partial z^2}-k^2g_{\vec{m}}^{(1,c)}(z,z_i)=0.
\end{equation}
\end{subequations}
The general solutions to these equations are as follows:
\begin{subequations}
\label{eqnset:general solutions}
\begin{equation}
\label{eqn:general solutions 1}
    g_{\vec{m}}^{(0)}(z,z_i)=A^{(0)}e^{-k(z-h_0)}+B^{(0)}e^{k(z-h_0)},  
\end{equation}
\begin{equation}
\label{eqn:general solutions 2}
    g_{\vec{m}}^{(1,s)}(z,z_i)=A^{(1,s)}e^{-\kappa (z-h_0)}+B^{(1,s)}e^{\kappa (z-h_0)},
\end{equation}
and
\begin{equation}
\label{eqn:general solutions 3}
    g_{\vec{m}}^{(1,c)}(z,z_i)=A^{(1,c)}e^{-k(z-h_0)}+B^{(1,c)}e^{k(z-h_0)}.
\end{equation}
\end{subequations}
where the constants A and B are determined by the boundary conditions. We additionally note that we have transformed $z$ to $z-h_0$ to reflect the unperturbed case of a particle above a surface with local deformation $h_0$. The exact constants and implementation details for many-particle simulations can be found in the \hyperref[section: Appendix]{Appendix}.
\section{Computational Methods}
For all self-energy calculations, particles are assumed to be to the right of the membrane to simplify calculations. The self-energy of such particles is then set equal to
\begin{equation}
    \label{eqn:self energy}
    U_{self}=\frac{l_B\epsilon_1Q_i^2}{2\sigma L_xL_y} \left(\sum_{\vec{m}'}\left\{A_{IV}^{\prime(0)}e^{-2k(z_i-h_0)} + h_cA_{IV}^{\prime(1,c)}e^{-2k (z_i-h_0)}\right\}+h_s\sum_{\vec{m}''}A_{IV}^{\prime(1,s)}e^{-(k+\kappa) (z_i-h_0)}\right),
\end{equation}
where $h$ is given in Equation \ref{eqn:h definition}, $l_B$ is the Bjerrum length, $\sigma$ is the particle diameter, the primed constants are equal to the A and B constants with the dependence on $z_i$ and $Q$ removed, the primed summation implies a sum over all $\vec{m}$ except $\vec{m}=0$, and the doubly primed summation implies a sum over all $\vec{m}$ except $\vec{m}=0$ and $\vec{m}+\vec{q}=0$. For the purposes of all self-energy experiments, the Bjerrum length is set to $0.7nm$ to approximate the behavior of particles in water and the ionic diameter is set to $0.35nm$. For this system, the unit of length used for all calculations is the particle diameter $\sigma$, but results are reported in the more universally reported Bjerrum length. For the purposes of the self-energy calculations, $L_x$ and $L_y$ are set to $5l_B$, $q_x$ is set to $2\pi / L_x$, $q_y$ is set to zero, the valence of the ions is set to one, $\epsilon_1$ is set to 80 (to mimic water), $\epsilon_2$ is set to 2 (to mimic a phospholipid bilayer), and the intermembrane distance is set to 7$l_B$.\supercite{dilger_dielectric_1979} A deformation amplitude $H$ of $0.25l_B$ is employed, and summations over $m_x$ and $m_y$ are performed over the interval $[-10,10]$ excluding zero.\par
In addition to single particle self-energy calculations, we also perform larger-scale Langevin dynamics simulations to elucidate the effects of dielectric mismatch coupled with surface curvature on systems of many charged particles. These simulations are performed using the velocity-Verlet algorithm with a timestep of $0.005d\sigma(m\beta)^{1/2}$.\supercites{telles_effects_2023}{paterlini_constant_1998} The box dimensions of the system are $5\times5\times30 l_B$; this box is periodic in the $x$ and $y$ directions, while the boundaries in the $z$ direction are treated with reflective boundary conditions. Each simulation box contains 18 positive ions with a valence of $+2$ and 36 negative ions with a valence of $-1$. These ions are evenly divided between the right and left sides of the membrane. The concentration of ions is calculated to be approximately 455 mmol; such a high salt concentration is relevant to nanoconfined systems. Simulations are run with and without dielectric mismatch; in the case of dielectric mismatch $\epsilon_1$ is set to 80 and $\epsilon_2$ is set to 2 to once again mimic a lipid bilayer in water. Since we employ an undulation in one dimension only, $q_y=0$ and we define $\lambda=1/q_x$ based on Equation \ref{eqn:h definition}. Simulations are conducted with surface undulation wavelength $\lambda$ equal to $5l_B$ and $2.5l_B$. The force due to dielectric mismatch is determined using the equations discussed in the \hyperref[section: Appendix]{Appendix} section \ref{subsection:implementation details}, while the standard electrostatic contribution is calculated using Coulomb potentials for short-range interactions and a slab-corrected Ewald summation for long-range interactions\supercite{santos_simulations_2016}. Other particle-particle interactions are calculated via a truncated Lennard-Jones potential using a cutoff of $2^{1/6}\sigma$. Each simulation run consists of $5\times10^4$ equilibration timesteps followed by $5\times10^5$ production timesteps with samples collected every 100 timesteps. 100 independent runs are performed for each set of parameters studied. Density profiles in $x$ and $z$ are calculated using bins of length $0.025l_B$ in the $x$ direction and $0.03l_B$ in the $z$ direction. All other parameters are identical to those used in the self-energy calculations.\par
\section{Results and Discussion}
\subsection{Self-Energy of a Single Particle}
Examining the self interaction energy of single particles near a bent dielectric interface is a valuable litmus test for our solution to the bent membrane problem. Near a dielectric interface, charged particles induce some surface charge at the dielectric boundary. For the water-lipid interface we are studying, the induced charge has the same sign as the charge on the particle and repels it; this can be verified with the method of images.\supercite{landau_electrodynamics_2009} As has been shown in the literature, the electrostatic force on a particle near a rough surface changes for particles located above surface deformations\supercite{solis_pimples_2021}.  A positive surface deformation moves the dielectric surface closer to any charged particle located above it. As a result, a particle near a positive surface deformation will feel a stronger repulsive force.\par
In addition to changing the distance between the particle and the surface, surface curvature itself alters the interfacial interaction of charged particles. The induced charge at a dielectric interface is proportional to $(\nabla\phi)\cdot \hat{n}$, where $\hat{n}$ is the unit vector normal to the interface. Given a particle above a convex surface, the angle between $\nabla\phi$ and $\hat{n}$ tends to be larger over the surface, causing there to be less charge accumulated there. The opposite is predicted for a particle above a concave surface.\supercite{solis_pimples_2021} Both this curvature effect and the surface distance effect discussed in the previous paragraph must be accounted for to estimate the interaction between a charged particle and a rough interface. A particle above a surface peak will have an increased self-energy due to the decreased distance to the peak surface relative to the undeformed membrane, but will have a decreased self-energy due to the increased convexity of surface at the peak. This trade-off is inverted for a particle above a surface trough.\par
\begin{figure*}[htbp]
    \centering
    \includegraphics[width=\textwidth]{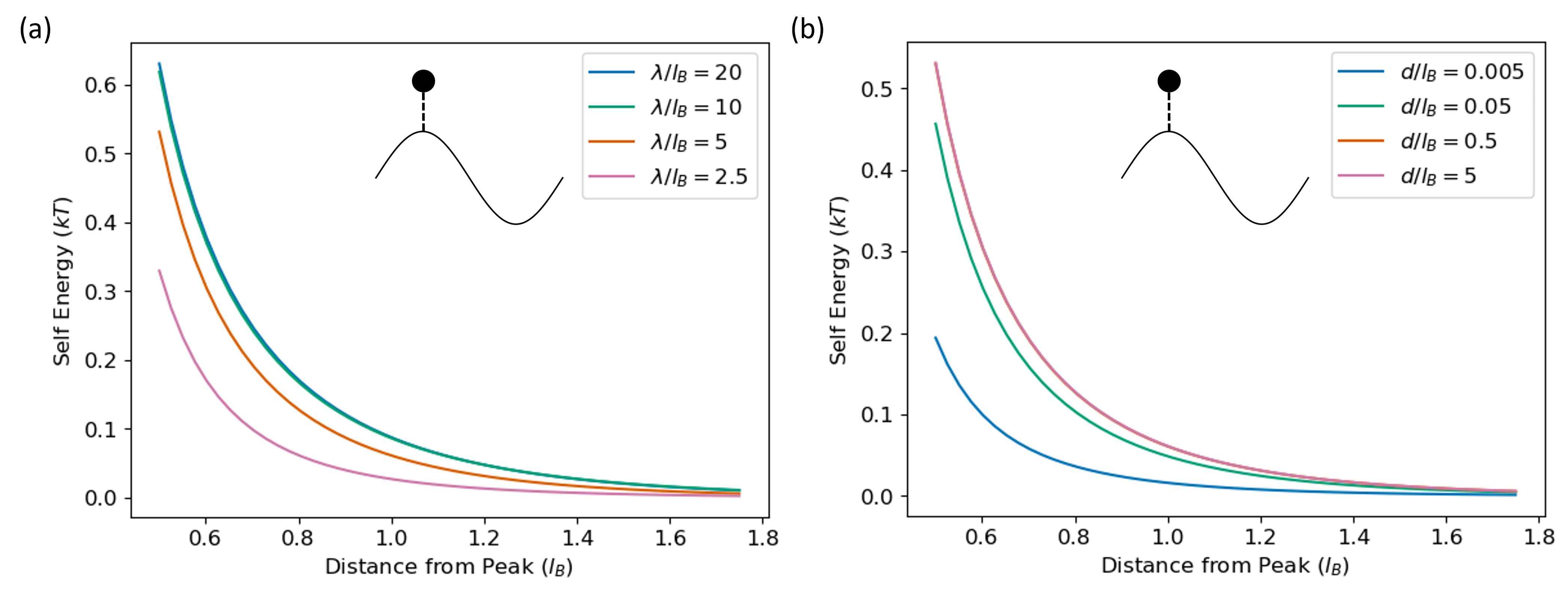}
    \caption{(a) Self-energy of a monovalent charged particle above a peak vs its $z$ position above a dielectric interface at varying wavelengths of interfacial oscillation. Results are depicted for $\lambda=2.5l_B$, $\lambda=5l_B$, $\lambda=10l_B$, and $\lambda=20l_B$. Increasing the wavelength increases the self-energy of charges above peaks due to decreased peak convexity at higher wavelengths. At high wavelengths, curvature effects become small and contribute little to changing the self-energy above the surface. (b) Self-energy of a charged particle above a peak vs its $z$ position, plotted for various values of the intermembrane distance. Results are depicted for intermembrane distances of $d=0.005l_B$, $d=0.05l_B$,$d=0.5l_B$,and $d=5l_B$. Smaller intermembrane distances decrease particle self-energy due to interactions with charges induced on the opposite face of the membrane; this effect vanishes at large intermembrane distances, presumably due to a vanishing self-energy contribution from charge induced on the opposite side of the membrane. The insets highlight the particle location relative to the curved membrane surface with deformation amplitude $H=0.25l_B$.}
    \label{fig:self energy vs freq dist}
\end{figure*}

We evaluate our method first by examining the energy of a single particle interacting with the charge it has induced at a dielectric interface. We first explore the magnitude of curvature effects on the self-energy. To accomplish this, we examine the effects of the oscillation wavelength $\lambda$ on the self-interaction energy for a particle above a peak; we calculate the wavelength of the oscillation from Equation \ref{eqn:h definition}. These results are shown in Figure \ref{fig:self energy vs freq dist}. Since decreasing surface convexity increases particle self-energy, we would expect increasing oscillation wavelengths to decrease peak convexity and therefore increase particle self-energy above peaks. This is exactly what we observe. At larger wavelengths, the contribution from peak convexity diminishes and eventually becomes negligible. At small wavelengths, however, the convexity has a large influence on self-energy, suggesting the importance of surface curvature in electrolyte structure of highly deformed charged systems.\par
In addition to undulation wavelength effects, we explore the consequences of manipulating the intermembrane distance on particle self-energy. We study this by calculating self-energies for a particle above an interfacial peak while varying the intermembrane distance. We achieve this by changing the parameter $d$ which is equal to half the intermembrane distance. As shown in Figure \ref{fig:self energy vs freq dist}, increases in intermembrane distance are associated with significant increases in self-energy, but these effects diminish quickly - the self-energies for the $d=0.5l_B$ and $d=5l_B$ cases are nearly identical. This indicates that intermembrane distance effects matter only at small separations. The self-energy contribution from charge on the opposite interface diminishes rapidly for thicker membranes, which agrees with the exponential scaling of polarization potential terms derived in the \ref{section:model} section. The contribution of induced charge at the opposite interface could likely be increased by reducing the dielectric constant inside the membrane, rendering intermembrane distance relevant for a wider array of length scales. Peristaltic deformations could also make this effect more relevant for surface troughs in particular by reducing the intermembrane distance at these locations. For the parameters studied, however, the effects of charge induced on the opposite interface play a significant role only at small intermembrane distances.\par
\begin{figure*}[htbp]
    \centering
    \includegraphics[width=\textwidth]{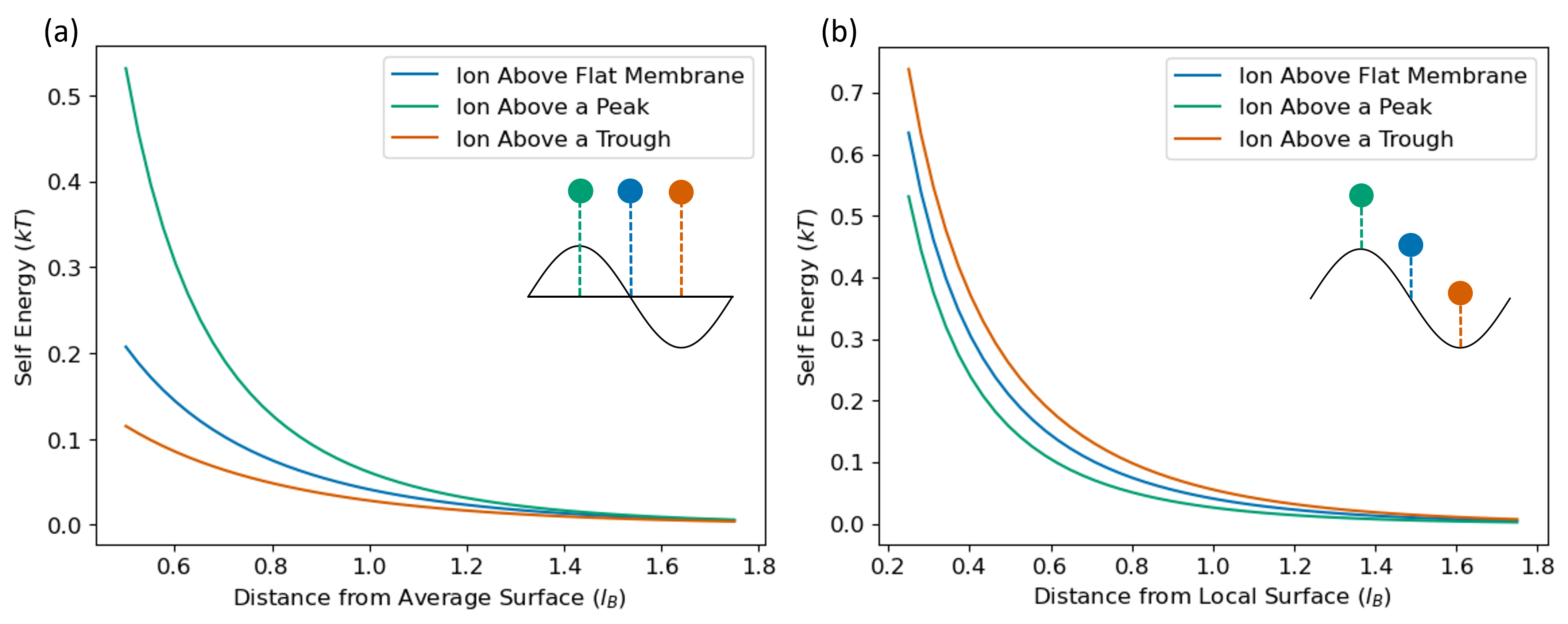}
    \caption{(a) Self-energy of a monovalent charged particle vs its distance from the \textit{unperturbed} membrane surface. Results are indicated for a charge directly above a peak (green), a charge above an undeformed section of membrane (blue), and a charge directly above a trough (orange). The self-energy of a charged particle is greatest above membrane peaks due to the associated reduction in distance between the particle and its induced charge. The inset highlights particle locations and distances measured to the undeformed surface with $\lambda=5l_B$ and $H=0.25l_B$. (b) Self-energy of a charged particle vs its distance from the local \textit{perturbed} surface. Results are indicated for a charge directly above a peak (green), a charge above an undeformed membrane element (blue), and a charge directly above a trough (orange). When maintaining a constant distance from the induced surface charge, the self-energy of a charged particle is greatest above membrane troughs due to their concavity. These results suggest a local depletion of charged particles from membrane troughs relative to peaks. The inset highlights particle locations and distances measured to the deformed membrane surface with $\lambda=5l_B$ and $H=0.25l_B$.}
    \label{fig:self energy vs z}
\end{figure*}

To further examine surface roughness effects, we first calculate the self-energy of particles as a function of their distance from the unperturbed surface, plotting simultaneously the self-energies for particles above surface peaks, surface troughs, and above undeformed surface elements. As shown in Figure \ref{fig:self energy vs z}, particles above surface peaks have higher self-energies than those above an undeformed part of the surface. The opposite holds true for particles above surface troughs. This is explicable via changes in surface-particle distance: for a given distance from the unperturbed surface, particles above peaks are closer to the bent surface than those above troughs, leading to an increased self-energy. These findings agree with recent literature studying electrostatic effects of bent dielectric interfaces.\supercite{solis_pimples_2021} The results confirm the role of surface-particle distance effects in self-energies near curved dielectrics. However, using distance from the unperturbed surface as the measure of distance obscures self-energy differences between particles at similar distances from the local deformed surface. In this case it is more revealing to compare self-energies between particles the same distance from the \textit{perturbed} surface than the \textit{unperturbed} one. These results are also shown in Figure \ref{fig:self energy vs z}.\par
When distance is measured to the local membrane surface, we find that particles above troughs have a higher self-energy than particles above peaks or undeformed portions of the membrane. In this case, particles above peaks have the lowest self-energies. These findings speak to the effects of surface curvature; since all particles are the same distance from their respective induced surface charges, the convexity or concavity of the local surface becomes the dominant factor in determining their relative self-energies. It is therefore reasonable to conclude that the concavity of the trough and convexity of the peak lead to respective increases and decreases in the self-energy of particles as described in previous work\supercites{goldstein_electric_1990}{solis_pimples_2021}. Since the surface curvature determines the energetic asymmetry between peaks and troughs, increasing membrane undulation should increase this asymmetry and the relative preference of charged particles for surface peaks over troughs.\par

\subsection{Simulations of Many-Particle Systems}
 Even given self-energy asymmetries between peaks and troughs, interactions are symmetric for electrolytes of ions with identical valences and radii. Therefore, there can be no net charge at the interface without introducing asymmetric electrolytes. While single-particle computations permit tractable analytical results, real systems are far more complicated, often consisting of many ions interacting with one another. To study these complex behaviors, we turn to Langevin dynamics simulations of 2:1 electrolytes with divalent positive and monovalent negative ions. It should be noted that our approach neglects the structural effects of particular solvents as has been shown to be important in the description of ion-adsorption at air-water interfaces\supercite{otten_elucidating_2012}; we believe that these effects can be important, but that capturing the dielectric effects at these interfaces can still capture much of the essential physics happening in their vicinity.\par
 To isolate the effects of asymmetric ion valence, we assume the ions have identical radii. First, we study the effects of dielectric mismatch near a curved membrane with an undulation wavelength of $5l_B$. We simulate this system both with and without dielectric mismatch. To elicit the effects of dielectric mismatch, we compare the density profiles of the ions between the two cases.\par
\begin{figure*}[htbp]
    \centering
    \includegraphics[width=\textwidth]{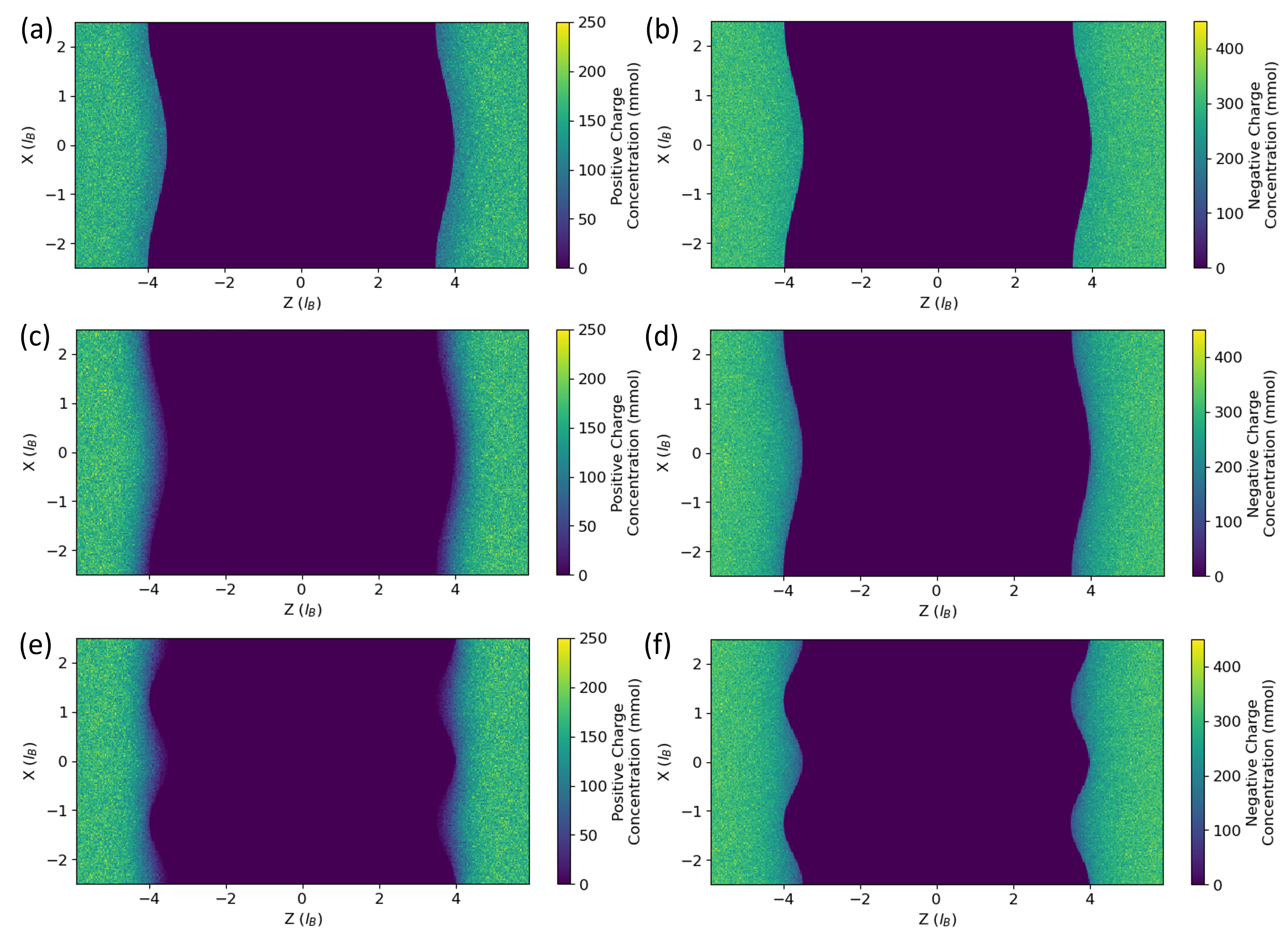}
    \caption{Charge densities of 2:1 electrolytes consisting of divalent positive and monovalent negative ions near a sinusoidally deformed dielectric interface. Subplots (a)-(d) depict a system with $\lambda=5l_B$ and amplitude $H=0.25l_B$ (calculated from Equation \ref{eqn:h definition}), where (a) and (b) depict a system without dielectric mismatch and (c) and (d) depict a system with dielectric mismatch.  Subplots (e) and (f) depict a system with dielectric mismatch, $\lambda=2.5l_B$, and $H=0.25l_B$. For both positive and negative ions, the incorporation of dielectric mismatch decreases charge density near the surface; this effect is stronger for the divalent positive charges than for the monovalent negative charges. This asymmetry between the divalent positive charges and monovalent negative charges is increased by decreasing $\lambda$ from $5l_B$ to $2.5l_B$. Surface charge depletion occurs preferentially from troughs over peaks in subplots (e) and (f); we believe this to be due to trough concavity and peak convexity increasing and decreasing the induced surface charge respectively.}
    \label{fig:charge densities lambda 10}
\end{figure*}
In Figure \ref{fig:charge densities lambda 10}, it is evident that the incorporation of dielectric mismatch leads to increased repulsion of ions from the surface; this effect occurs on both sides of the membrane. For the 2:1 system, the repulsion is stronger for the positive divalent ions than for the negative monovalent ions, which is expected since the strength of induced charge repulsion scales with the square of the ion valence. These results assume identically-sized ions although typically cations are smaller than anions. It should be noted that even without dielectric mismatch, there is some local depletion of ions from the interface: this is explicable by the fact that ions in solution experience a net attraction to ions in the bulk. These results agree with those calculated by Wu et al. using numerical boundary element methods, indicating that our method reproduces all of the essential effects associated with rough dielectric interfaces.\supercite{wu_asymmetric_2018}\par
According to Figure \ref{fig:self energy vs freq dist}, curvature effects become more potent at lower surface undulation wavelengths. The potential to use surface geometry to control ion positions near these curved dielectric interfaces motivates another set of simulations examining a membrane with an undulation wavelength of $2.5l_B$. This smaller wavelength implies an increased importance of peak and trough curvature. By increasing the difference in curvature between peaks and troughs, we hypothesize that we can more strictly control ionic concentrations near the interface. These changes in ion adsorption are likely related to an electrostatic contribution to the surface tension of the membrane and by extension its distribution of height-height correlations.\supercite{galib_mass_2017} By sampling a variety of deformation wavelengths, one could estimate a capillary wavelength for the system driven by changes in ion behavior near the interface. This analysis reaches beyond the scope of the current work but is well worth exploring in the future.\par
\begin{figure*}[htbp]
    \centering
    \includegraphics[width=\textwidth]{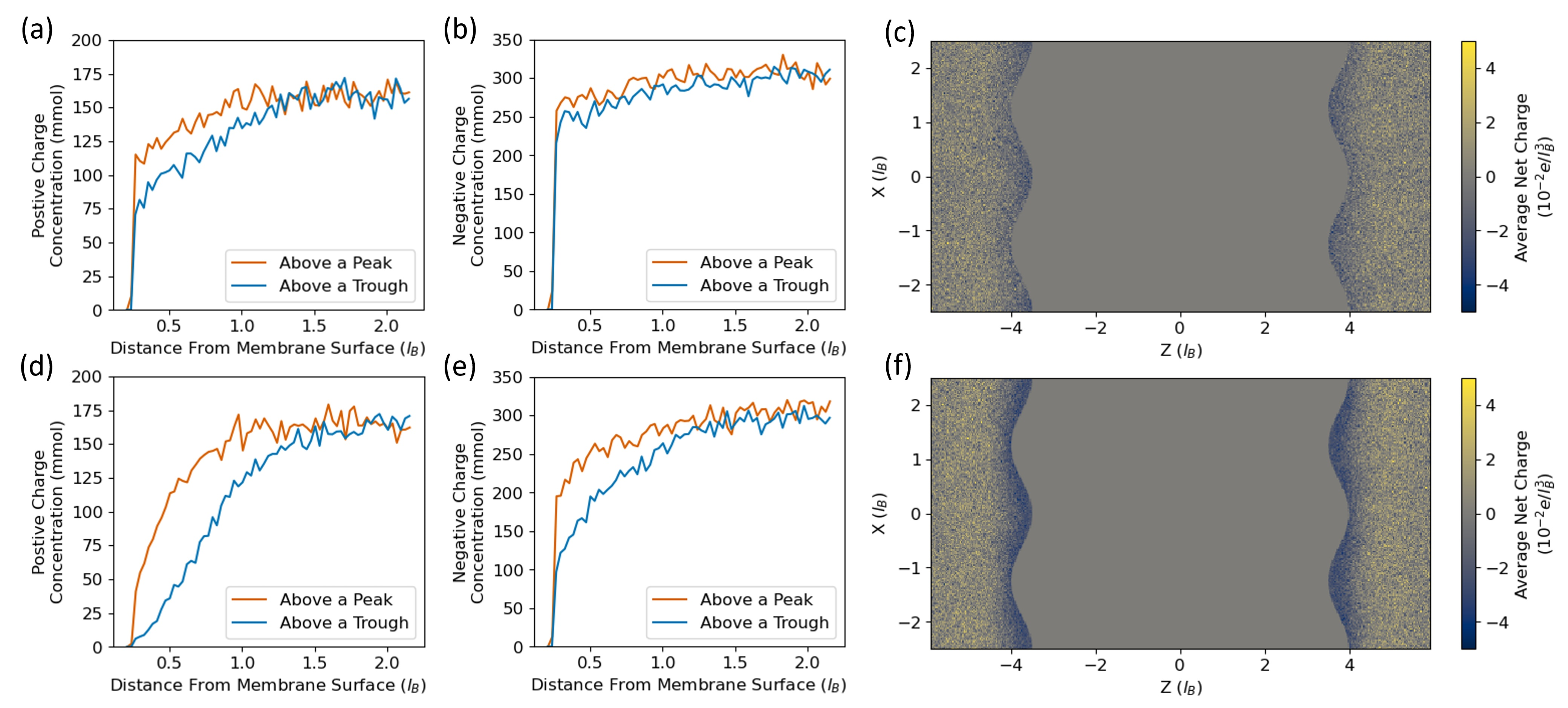}
    \caption{Charge densities of divalent positive and monovalent negative ions near a sinusoidally deformed dielectric surface with $\lambda=2.5l_B$. Top: (a) Positive and (b) negative charge densities above peaks (orange) and troughs (blue) of a deformed surface without dielectric mismatch. (c) Two-dimensional plot of the net charge density near the surface. Bottom: (d) positive and (e) negative charge densities above peaks (orange) and troughs (blue) of a deformed surface with dielectric mismatch. (f) Two-dimensional plot of the net charge density near the surface of a highly deformed dielectric interface. Some preference for surface peaks over troughs is visible for the system without dielectric mismatch; this can be explained by increased distance between troughs from the ionic bulk. For the system with dielectric mismatch, increased surface curvature causes a strong depletion of ions from surface troughs relative to peaks. This depletion is exaggerated for the positive ions due to their higher valence. The stronger depletion of divalent ions from troughs leads to net charge patterning in accordance with the interfacial geometry.}
    \label{fig:net charge lambda 2.5}
\end{figure*}

The results show that, as proposed, asymmetric dielectric effects become much stronger for a more oscillatory membrane. As shown in Figure \ref{fig:charge densities lambda 10} (e), positive charges are even more strongly repelled from troughs due to the effects of curvature on polarization forces. This confirms the hypothesis that increased magnitudes of membrane curvature lead to more pronounced asymmetry between peak and trough surface charge densities. In addition to the density maps in Figure \ref{fig:charge densities lambda 10}, we explicitly calculate the charge densities of the ionic species above peaks and troughs respectively to quantify any differences between them. We also calculate the net charge density distribution near the interface. These results are displayed in Figure \ref{fig:net charge lambda 2.5}.\par
In Figure \ref{fig:net charge lambda 2.5} (d) and (e), we observe that incorporating dielectric mismatch leads to reduced repulsion for ions above peaks compared to those above troughs. This preference is displayed to a lesser extent in the system without dielectric mismatch, which could be due to the fact that the divalent ions near troughs are further from the ionic bulk than those near peaks. Based on the increase in the effect for the increased ion valency, this appears to be an ionic correlation effect. Correlations arising due to lateral repulsion between multivalent ions adsorbed to strongly oppositely charged surfaces generate two-dimensional patterns known as Wigner crystals.\supercite{rouzina_macroion_1996} These Wigner crystals are modified by dielectric mismatch.\supercite{samaj_ground-state_2012}. While the effect is shown to be negligible for neutral surfaces,\supercite{agrawal_nature_2023} it is possible that some smaller correlations arise in multivalent salts. Indeed, while this preference exists without dielectric mismatch, it is greatly enhanced by its incorporation. By comparing Figure \ref{fig:net charge lambda 2.5} (c) and (f), we observe that the magnitude of the net charge near surface troughs increases substantially for a surface with dielectric mismatch. This is due to excess depletion of divalent positive charges from troughs and demonstrates that interfacial charge patterning in asymmetric ionic systems is greatly enhanced by the combination of curvature and dielectric effects. These results show that simply by changing the curvature of a dielectric interface, it is possible to create a wide variety of different surface charge patterns. This promises an exciting array of applications in bioinspired and iontronic materials.
\section{Conclusions}
To determine the effects of dielectric mismatch near curved membranes, we introduce a perturbative formulation for calculating electrostatic potentials from particles near rough interfaces in slab geometries. This formulation uses a first-order approximation to estimate electrostatic potentials stemming from individual particles; linear combinations of these potentials yield results for systems containing many charges. Although we only consider sinusoidal deformations here, this perturbative scheme can account for general deformations by combining sinusoidal perturbations in two dimensions. This approach allows for the study of arbitrarily complex curved interfaces. This can be incorporated into a Langevin approach to study the effects of ionic size and valence.\par
Examining the self-energies of individual particles, we find that particles feel two effects from the deformation of dielectric surfaces. The first is related to changes in distance between a particle and the surface, while the second is related to the concavity or convexity of the local surface. Taken together, these phenomena cause troughs to repel particles near the surface more strongly than peaks. We show that the effects of curvature become more potent with decreasing undulation wavelength of the interface which is accompanied by increasing the magnitude of the interfacial curvature. We also examine effects of the intermembrane distance on the self-energy, concluding that opposite interfaces only interact meaningfully with one another if the intermembrane distance is very small.\par
Simulations of highly concentrated electrolyte solutions employing Langevin dynamics show that these single particle self-energy concepts are broadly applicable to larger charged systems. Our Langevin approach permits incorporation of hard-core repulsion and multi-particle correlation effects, which is well suited to studying the highly concentrated electrolytes relevant to confined systems. The addition of dielectric mismatch leads to increased depletion of ions at membrane surfaces, occurring preferentially at troughs due to their concavity. By employing a 2:1 asymmetric electrolyte with divalent positive and monovalent negative ions, we can selectively generate net charge densities via preferential depletion of positive ions near surface troughs. These results present the possibility to create charge patterns at dielectric boundaries by tuning interfacial geometry.\par
In this work, we show that particles can be selectively driven from different interfacial locations depending on surface curvature. Given the ability to tune interfacial roughness in electronics and other devices, this directly leads to the ability to selectively deplete charged particles from interfaces. The effect could be especially potent in the presence of multivalent ions, where induced charge effects are stronger. By altering only boundary conditions, our method can be extended to study ion distributions in electrolytes confined by rough interfaces, offering immense potential to the growing field of iontronics. It provides a fast, robust way to validate numerical methods and establish appropriate parametrizations for interfacial ionic control, rendering it invaluable to the nanoscience of charged systems.\par
\section{Supplementary Material}
The supplementary material includes graphs displaying the full range of particle self-energy with varying lateral displacement near a dielectric interface, the charges directly above peaks and troughs for the Langevin system with $\lambda=5l_B$, charge density maps for the Langevin system with $\lambda=2.5l_B$ and no dielectric mismatch, simulation results from a system with $\epsilon_1=80$ and $\epsilon_2=20$, and validation of the approach described in this work using numerical methods.
\section{Acknowledgements}
The authors acknowledge support from the Department of Energy, Office of Basic Energy Sciences under contract DE-FG02-08ER46539. N.P. acknowledges support from the Ryan Fellowship through the International Institute for Nanotechnology at Northwestern University and additionally the support of the Hierarchical Materials Cluster Program at Northwestern University. A. P. d.S. thanks the Department of Materials Science and Engineering at Northwestern University for appointing him Eschbach Visiting Professor. A. P. d.S. also thanks FAPERGS and CNPq. M. O. d.l.C. research was supported in part by grant NSF PHY-1748958 to the Kavli Institute for Theoretical Physics (KITP).
\section{Conflict of Interest Statement}
The authors have no conflicts to disclose.
\section{Author Contributions}
\textbf{N. P.}: Conceptualization (equal); formal analysis (lead); investigation (lead); methodology (lead); software (equal); writing - original draft (lead); writing - review and editing (equal). \textbf{A. P. d.S.}: Formal analysis (supporting); investigation (supporting); methodology (supporting); software (equal); writing - review and editing (equal). \textbf{A. E.}: Formal analysis (supporting); software (equal) \textbf{M. O. d.l.C.}: Conceptualization (equal); formal analysis (supporting); investigation (supporting); funding acquisition (lead); supervision (lead); project administration (lead); writing - review and editing (equal).
\section{Data Availability Statement}
The data that support the findings of this study are available within the article [and its supplementary material].
\newcounter{Aequ}
\newenvironment{AEquation}
  {\stepcounter{Aequ}%
    \addtocounter{equation}{-1}%
    \renewcommand\theequation{A\arabic{Aequ}}\equation}
  {\endequation}
\section{Appendix: Exact Solutions and Implementation}
\label{section: Appendix}
We solve Equation \ref{eqn:ansatz plus substitution full} by solving each term in the summations over $\vec{m}$ separately. Moreover, we solve independently the terms proportional to $h^0$, $h_s^1$, and $h_c^1$. This leads to the following three equations:
\begin{subequations}
\begin{AEquation}
    \label{eqn:zeroth order equation SI}
    \frac{\partial^2 g_{\vec{m}}^{(0)}(z,z_i)}{\partial z^2}-k^2g_{\vec{m}}^{(0)}(z,z_i)=-\frac{4 \pi Q_i}{\epsilon_1L_xL_y} \delta (z-z_i),
\end{AEquation}
\begin{AEquation}
    \label{eqn:first order equation SI}
    \frac{\partial^2 g_{\vec{m}}^{(1,s)}(z,z_i)}{\partial z^2}-\kappa^2g_{\vec{m}}^{(1,s)}(z,z_i)=0,
\end{AEquation}and
\begin{AEquation}
    \label{eqn:first order equation c term SI}
    \frac{\partial^2 g_{\vec{m}}^{(1,c)}(z,z_i)}{\partial z^2}-k^2g_{\vec{m}}^{(1,c)}(z,z_i)=0.
\end{AEquation}
\end{subequations}
We note that Equation \ref{eqn:first order equation c term SI} simply reflects the $\vec{q}=0$ case of Equation \ref{eqn:first order equation SI} and will be treated as such in the following derivations for brevity.
\subsection{Case 1: General Case}
\label{subsection:general case solutions}
In the general case, the solution to the zeroth order equation (Equation \ref{eqn:zeroth order equation SI}) can be written:
\begin{AEquation}
    \label{eqn:general case 0 order solution}
    g_{\vec{m}}^{(0)}(z,z_i)=A^{(0)}e^{-k(z-h_0)}+B^{(0)}e^{k(z-h_0)},
\end{AEquation}
and solution to the first order sinusoidal Equation (Equation \ref{eqn:first order equation SI}) is:
\begin{AEquation}
    \label{eqn:general case 1 order solution}
    g_{\vec{m}}^{(1,s)}(z,z_i)=A^{(1,s)}e^{-\kappa (z-h_0)}+B^{(1,s)}e^{\kappa (z-h_0)},
\end{AEquation}
where A and B are constants determined by the boundary conditions and $z$ is transformed to $z-h_0$ to represent the distance from the center of a deformed membrane. It is immediately evident that the solution to the first order constant equation is, as mentioned above, identical to Equation \ref{eqn:general case 1 order solution} with $\vec{q}$ set equal to zero. In this case $\kappa=k$.\par
To apply the boundary conditions, we divide our system into four regions as shown in Figure \ref{fig:box with regions}.
We apply boundary conditions mandating that the potentials $\phi$, as well as $\nabla\phi\cdot\hat{n}$, in different regions be equal at the region boundaries. Here, $\hat{n}$ represents the unit normal vector to the region boundary. This reduces to $\hat{n}=(0,0,1)$ for a flat boundary; for a general bounding surface $z=f(x,y)$ it takes the form $\hat{n}=(-\frac{\partial f}{\partial x},-\frac{\partial f}{\partial y},1)$. The one exception to this is at the boundary between Regions III and IV, where the discontinuity in the derivative of the potential across the interface is set equal to $\frac{4 \pi Q_i}{\epsilon_1}$ due to the delta function in Equation \ref{eqn:zeroth order equation SI}. This leaves us with the following six equations:
\begin{AEquation}
\begin{multlined}
    \label{eqn:boundary conditions}
     \\ \mathbf{I.} \phi_{I}(-d+h)=\phi_{II}(-d+h), \\
    \mathbf{II.} \phi_{II}(d+h)=\phi_{III}(d+h), \\
    \mathbf{III.} \phi_{III}(z_i)=\phi_{IV}(z_i), \\ 
    \mathbf{IV.} \epsilon_1(\nabla \cdot \phi_{I})|_{z=-d+h} = \epsilon_2(\nabla \cdot \phi_{II})|_{z=-d+h}, \\
    \mathbf{V.} \epsilon_2(\nabla \cdot \phi_{II})|_{z=d+h} = \epsilon_1(\nabla \cdot \phi_{III})|_{z=d+h}, \\
    \mathbf{VI.} \frac{\partial \phi_{III}}{\partial z}\bigg|_{z=z_i} - \frac{\partial \phi_{IV}}{\partial z}\bigg|_{z=z_i} = \frac{4 \pi Q_i}{\epsilon_1},
\end{multlined}
\end{AEquation}
where the subscripts indicate which region the potential is describing. Expanding each term to first order and accounting for the fact that the potential must approach zero as $z \rightarrow \pm \infty$, we obtain the following system of equations:
\begin{AEquation}
    \begin{multlined}
        \label{eqn:boundary conditions expanded}
        \\ \mathbf{I.} B^{(0)}_{I}e^{-kd}(1+kh)+hB^{(1)}_{I}e^{-kd}= A^{(0)}_{II}e^{kd}(1-kh)+B^{(0)}_{II}e^{-kd}(1+kh)+h(A^{(1)}_{II}e^{kd}+B^{(1)}_{II}e^{-kd}), \\
    \mathbf{II.} A^{(0)}_{II}e^{-kd}(1-kh)+B^{(0)}_{II}e^{kd}(1+kh)+h(A^{(1)}_{II}e^{-kd}+B^{(1)}_{II}e^{kd})= \\ A^{(0)}_{III}e^{-kd}(1-kh)+B^{(0)}_{III}e^{kd}(1+kh)+h(A^{(1)}_{III}e^{-kd}+B^{(1)}_{III}e^{kd}), \\
    \mathbf{III.} A^{(0)}_{III}e^{-kz_i}+B^{(0)}_{III}e^{kz_i}+h(A^{(1)}_{III}e^{-kz_i}+B^{(1)}_{III}e^{kz_i}) =A^{(0)}_{IV}e^{-kz_i}+hA^{(1)}_{IV}e^{-kz_i}, \\ 
    \mathbf{IV.} \epsilon_1(kB^{(0)}_{I}e^{-kd}(1+kh)+h(kB^{(1)}_{I}e^{-kd}-\frac{\kappa^2-k^2-q^2}{2}B^{(0)}_{I}e^{-kd}))= \\ \epsilon_2(-kA^{(0)}_{II}e^{kd}(1-kh)+kB^{(0)}_{II}e^{-kd}(1+kh)+h(-kA^{(1)}_{II}e^{kd}+kB^{(1)}_{II}e^{-kd}-\\ \frac{\kappa^2-k^2-q^2}{2}(A^{(0)}_{II}e^{kd}+B^{(0)}_{II}e^{-kd}))), \\
    \mathbf{V.} \epsilon_2(-kA^{(0)}_{II}e^{-kd}(1-kh)+kB^{(0)}_{II}e^{kd}(1+kh)+h(-kA^{(1)}_{II}e^{-kd}+kB^{(1)}_{II}e^{kd}-\\ \frac{\kappa^2-k^2-q^2}{2}(A^{(0)}_{II}e^{-kd}+B^{(0)}_{II}e^{-kd})))= \\ \epsilon_2(-kA^{(0)}_{III}e^{-kd}(1-kh)+kB^{(0)}_{III}e^{kd}(1+kh)+h(-kA^{(1)}_{III}e^{-kd}+kB^{(1)}_{III}e^{kd}-\\ \frac{\kappa^2-k^2-q^2}{2}(A^{(0)}_{III}e^{-kd}+B^{(0)}_{III}e^{-kd}))), \\
    \mathbf{VI.} -kA^{(0)}_{III}e^{-kz_i}+kB^{(0)}_{III}e^{kz_i}+h(-kA^{(1)}_{III}e^{-kz_i}+kB^{(1)}_{III}e^{kz_i}) = kA^{(0)}_{IV}e^{-kz_i}-hkA^{(1)}_{IV}e^{-kz_i}, \\
    \end{multlined}
\end{AEquation}
where the superscripts indicate which order term (zeroth or first) each constant corresponds to and $q=\sqrt{q_x^2+q_y^2}$.\par
This system of equations can be easily solved by separating it into two sets of six equations each with one set containing terms of order $h^0$ and the other set containing terms of order $h^1$. Solving these two sets of equations, we find that the zero-order constants take the following values:
\begin{AEquation}
\begin{multlined}
    \label{eqn:values of zero order constants}
     \\ B^{(0)}_{I}= \frac{8e^{4dk-k(z_i-h_0)}\epsilon_2 \pi Q_i}{k(e^{4dk}(\epsilon_1+\epsilon_2)^2-(\epsilon_2-\epsilon_1)^2)}\\
    A^{(0)}_{II}= \frac{4e^{2dk-k(z_i-h_0)}(\epsilon_2-\epsilon_1)\pi Q_i}{k(e^{4dk}(\epsilon_1+\epsilon_2)^2-(\epsilon_2-\epsilon_1)^2)},\\
    B^{(0)}_{II}=\frac{4e^{2dk-k(z_i-h_0)}(\epsilon_2-\epsilon_1)\pi Q_i}{k(e^{4dk}(\epsilon_1+\epsilon_2)^2-(\epsilon_2-\epsilon_1)^2)}, \\
    A^{(0)}_{III}=\frac{2e^{2dk-k(z_i-h_0)}(-1+e^{4dk})(\epsilon_1-\epsilon_2)(\epsilon_1+\epsilon_2)\pi Q_i}{\epsilon_1 k(e^{4dk}(\epsilon_1+\epsilon_2)^2-(\epsilon_2-\epsilon_1)^2)}, \\
    B^{(0)}_{III}=\frac{2e^{-k(z_i-h_0)}\pi Q_i}{\epsilon_1 k}, \\
    A^{(0)}_{IV}=\frac{2e^{-k(z_i-h_0)}(-e^{2k(z_i-h_0)}(\epsilon_2-\epsilon_1)^2+e^{2dk}(\epsilon_2-\epsilon_1)(\epsilon_2+\epsilon_1)+e^{2k(2d+(z_i-h_0))(\epsilon_2+\epsilon_1)^2+e^{6dk}(\epsilon_1^2-\epsilon_2^2))\pi Q_i}}{\epsilon_1 k(e^{4dk}(\epsilon_1+\epsilon_2)^2-(\epsilon_2-\epsilon_1)^2)}, \\
\end{multlined}
\end{AEquation}
where we note that $z_i$ has been transformed to $z_i-h_0$ to represent a particle above a flat surface of distance $h_0$ above the unperturbed flat surface (where $h_0=0$). The first order constants arising from the sinusoidal perturbation $h_c$ (denoted by a superscript $(1,s)$) are
\begin{AEquation}
    \begin{multlined}
    \label{eqn:values of first order sinusoidal constants}
    \\B^{(1,s)}_{I}=-\frac{4F_0\epsilon_2 \sinh (d(k-\kappa))(F_{11}+F_{12})}{D},\\
    A^{(1,s)}_{II}= \frac{F_0(F_{21}+F_{22}+F_{23})}{2D}, \\
    B^{(1,s)}_{II}= -\frac{(F_0(F_{31}+F_{32}-F_{33})}{2D},\\
    A^{(1,s)}_{III}= -\frac{(F_0(F_{41}+F_{42}+F_{43}+F_{44})}{2\epsilon_1D},\\
    B^{(1,s)}_{III}= 0,\\
    A^{(1,s)}_{IV}= -\frac{(F_0(F_{41}+F_{42}+F_{43}+F_{44})}{2\epsilon_1D}.\\
\end{multlined}
\end{AEquation}
where we note that
\begin{AEquation}
    \begin{multlined}
        \label{defining some terms to make the first order expressions less messy}
        \\ F_0=4e^{d(k+\kappa)-k(z_i-h_0)}(\epsilon_2-\epsilon_1)\pi Q_i, \\
        F_{11}=(-2\epsilon_2 k \kappa+\epsilon_1(k^2+\kappa^2-q^2))\cosh (d(k+\kappa)), \\
        F_{12}=(-2\epsilon_1 k \kappa+\epsilon_2(k^2+\kappa^2-q^2))\sinh (d(k+\kappa)), \\
        F_{21}=2(1+e^{2d(k+\kappa)})\epsilon_2\epsilon_1(k+\kappa-q)(k+\kappa+q),\\
        F_{22}=\epsilon_1^2((-1+e^{4dk})k^2-2(1+e^{4dk}-2e^{2d(k+\kappa)})k\kappa+(-1+e^{4dk})(\kappa-q)(\kappa+q)),\\
        F_{23}=\epsilon_2^2(-((1+e^{4dk}-2e^{2d(k+\kappa)})k^2)+2(-1+e^{4dk})k\kappa-(1+e^{4dk}-2e^{2d(k+\kappa)})(\kappa-q)(\kappa+q)), \\
        F_{31}=e^{2d(2k+\kappa)}(\epsilon_2+\epsilon_1)^2(k-\kappa-q)(k-\kappa+q), \\
        F_{32} = e^{2d\kappa}(\epsilon_2-\epsilon_1)(\epsilon_2+\epsilon_1)(k+\kappa-q)(k+\kappa+q), \\
        F_{33} = 2e^{2dk}(\epsilon_2-\epsilon_1)(2\epsilon_1k\kappa+\epsilon_2(k^2+\kappa^2-q^2)), \\
        F_{41} = 2(-1+e^{4dk})(-1+e^{4d\kappa})\epsilon_2^3k\kappa, \\
        F_{42} = (-1+e^{4dk})(-1+e^{4d\kappa})\epsilon_1^3(k^2+\kappa^2-q^2),\\
        F_{43}=2\epsilon_2\epsilon_1^2((-1+e^{4d(k+\kappa)})k^2+(1+e^{4dk}+e^{4d\kappa}-4e^{2d(k+\kappa)}, \\ +e^{4d(k+\kappa)})k\kappa+(-1+e^{4d(k+\kappa)})(\kappa-q)(\kappa+q)), \\
        F_{44} = \epsilon_2^2\epsilon_1((1+e^{4dk}+e^{4d\kappa}-4e^{2d(k+\kappa)}+e^{4d(k+\kappa)})k^2+4(-1+e^{4d(k+\kappa)})k\kappa+ \\(1+e^{4dk}+e^{4d\kappa}-4e^{2d(k+\kappa)}+e^{4d(k+\kappa)})(\kappa-q)(\kappa+q))\\
        D=k\kappa (2\epsilon_2\epsilon_1 \cosh (2dk)+(\epsilon_2^2+\epsilon_1^2)\sinh (2dk))(2\epsilon_2\epsilon_1 \cosh (2d\kappa)+(\epsilon_2^2+\epsilon_1^2)\sinh (2d\kappa)). \\
    \end{multlined}
\end{AEquation}
The first order constants arising from the constant perturbation $h_c$ (denoted by a superscript $(1,c)$) can be obtained by setting $\vec{q}=0$ in the first order equations and are
\begin{AEquation}
    \begin{multlined}
    \label{eqn:values of first order constant constants}
    \\B^{(1,c)}_{I}=0, \\
    A^{(1,c)}_{II}=\frac{8e^{2dk-k(z_i-h_0)}(\epsilon_2-\epsilon_1)\pi Q_i}{-(\epsilon_2-\epsilon_1)^2+e^{4dk}(\epsilon_2+\epsilon_1)^2}, \\
    B^{(1,c)}_{II}=0, \\
    A^{(1,c)}_{III}=\frac{4e^{2dk-k(z_i-h_0)}(e^{4dk}-1)(\epsilon_1^2-\epsilon_2^2)\pi Q_i}{\epsilon_1(e^{4dk}(\epsilon_2+\epsilon_1)^2-(\epsilon_2-\epsilon_1)^2)}, \\
    B^{(1,c)}_{III}=0, \\
    A^{(1,c)}_{IV}=\frac{4e^{2dk-k(z_i-h_0)}(e^{4dk}-1)(\epsilon_1^2-\epsilon_2^2)\pi Q_i}{\epsilon_1(e^{4dk}(\epsilon_2+\epsilon_1)^2-(\epsilon_2-\epsilon_1)^2)}. \\
\end{multlined}
\end{AEquation}
We also highlight the fact that $A_{III}^{(1,s)}=A_{IV}^{(1,s)}$ and $A_{III}^{(1,c)}=A_{IV}^{(1,c)}$. For a system without any dielectric mismatch, we find the following:
\begin{AEquation}
    \begin{multlined}
        \label{eqn:terms without polarization}
        \\B_{I}^{(0)}=B_{II}^{(0)}=B_{III}^{(0)}= \frac{2e^{-k(z_i-h_0)}\pi Q_i}{\epsilon_1 k}, \\
        A_{IV}^{(0)}=\frac{2e^{k(z_i-h_0)}\pi Q_i}{\epsilon_1 k},\\
    \end{multlined}
\end{AEquation}
with all other terms being equal to 0 in this case. For calculation purposes, we then consider the potential as being divided into a standard electrostatic contribution and a polarization contribution. The standard electrostatic portion can be treated using versions of the Ewald method modified for slab geometries.\supercites{santos_simulations_2016}{yeh_ewald_1999} We separate out the contribution from polarization by subtracting the terms in Equation \ref{eqn:terms without polarization} from our Equation \ref{eqn:values of zero order constants}. This serves to modify the $B_{II}^{(0)}$,$B_{II}^{(0)}$,$B_{III}^{(0)}$,and $A_{IV}^{(0)}$ terms, which become:
\begin{AEquation}
    \begin{multlined}
        \label{eqn:polarization terms}
        \\B_{I}^{(0)}= -\frac{2e^{-k(z_i-h_0)}(-1+e^{4dk})(\epsilon_2-\epsilon_1)^2\pi Q_i}{\epsilon_1 k(e^{4dk}(\epsilon_2+\epsilon_1)^2-(\epsilon_2-\epsilon_1)^2)},\\
        B_{II}^{(0)}= \frac{2e^{-k(z_i-h_0)}(-\epsilon_2+\epsilon_1)(-\epsilon_2+\epsilon_1+e^{4dk}(\epsilon_2+\epsilon_1)) \pi Q_i}{\epsilon_1 k(e^{4dk}(\epsilon_2+\epsilon_1)^2-(\epsilon_2-\epsilon_1)^2)},\\
        B_{III}^{(0)}= 0, \\
        A_{IV}^{(0)}= \frac{2e^{2dk-k(z_i-h_0)}(-1+e^{4dk})(-\epsilon_2^2+\epsilon_1^2) \pi Q_i}{\epsilon_1 k(e^{4dk}(\epsilon_2+\epsilon_1)^2-(\epsilon_2-\epsilon_1)^2)}.\\
    \end{multlined}
\end{AEquation}
The process outlined in this section can be repeated with nearly identical results for a charge on the left side of the membrane. A complete solution considering charges on either side of the membrane is implemented as part of the Langevin dynamics simulations.\par
\subsection{\texorpdfstring{Case 2: $k=0$}{Case 2: k=0}}
In the $k=0$ case, there are divergent terms which must be accounted for in the $A$ and $B$ constants. Expanding the zeroth order $A$ and $B$ polarization energy terms around $k=0$, we obtain the following results:
\begin{AEquation}
    \begin{multlined}
        \label{eqn:k=0 zero order expansions}
        \\B_{I}^{(0)}=-\frac{2(d(\epsilon_2-\epsilon_1)^2 \pi Q_i)}{\epsilon_2\epsilon_1^2}+\mathcal{O}(k), \\
        A_{II}^{(0)}=\frac{(\epsilon_2-\epsilon_1) \pi Q_i}{\epsilon_2\epsilon_1 k}-\frac{(\epsilon_2-\epsilon_1) \pi Q_i (d\epsilon_2^2+d\epsilon_1^2+\epsilon_1\epsilon_2(z-h_0)+\epsilon_2\epsilon_1 (z_i-h_0))}{\epsilon_2^2\epsilon_1^2}+\mathcal{O}(k), \\
        B_{II}^{(0)}=\frac{(\epsilon_1-\epsilon_2) \pi Q_i}{\epsilon_2\epsilon_1 k}-\frac{(\epsilon_1-\epsilon_2) \pi Q_i (-d\epsilon_2^2+d\epsilon_1^2-\epsilon_1\epsilon_2(z-h_0)+\epsilon_2\epsilon_1 (z_i-h_0))}{\epsilon_2^2\epsilon_1^2}+\mathcal{O}(k), \\
        A_{III}^{(0)}=\frac{2d(\epsilon_1-\epsilon_2)(\epsilon_1+\epsilon_2)\pi Q_i}{\epsilon_2\epsilon_1^2}+\mathcal{O}(k), \\
        B_{III}^{(0)} = 0, \\
        A_{IV}^{(0)}=\frac{2d(\epsilon_1^2-\epsilon_2^2)\pi Q_i}{\epsilon_2\epsilon_1^2}+\mathcal{O}(k). \\   
    \end{multlined}
\end{AEquation}

It can be observed from Equation \ref{eqn:k=0 zero order expansions} that only the energetic terms corresponding to Region II (the region between the membranes) depend on $z$; therefore we need not consider the $k=0$ case for the purpose of calculating forces on the particles, which all lie in Regions I, III, and IV. Moreover, the energetic terms in Regions I, III, and IV are all constants to zero order in $k$. This will lead to an increase in the system energy by some constant which does not depend on the positions of the particles and can be safely renormalized away.\par
Redefining the first-order terms in the $k=0$ case merits an alternative approach. To calculate these terms, we simply solve the first-order piece of Equation \ref{eqn:boundary conditions expanded} while setting $k=0$. Solving this leads to all first order constants being equal to zero, meaning that they contribute nothing to the electrostatic energy or force. Pairing this with the result above (Equation \ref{eqn:k=0 zero order expansions}), the $k=0$ term may be excluded entirely from our sum over $\vec{m}$ in Equation \ref{eqn:ansatz}.
\subsection{\texorpdfstring{Case 3: $\kappa=0$}{Case 3: kappa=0}}
Much like the $k=0$ case, the nature of our expressions for the constants $A$ and $B$ lead to divergences at $\kappa=0$. Since the zeroth order terms and first order constant terms are not proportional to $\kappa$ in any way, they are not considered here. The first order sinusoidal terms, however, must be carefully examined. To do so, we expand these terms around $\kappa=0$, leading to the following:

\begin{AEquation}
    \begin{multlined}
        \label{eqn:kappa=0 expansions for first order terms}
        \\B_{I}^{(1,s)}=-4\epsilon_2(\epsilon_2-\epsilon_1)\pi Q_i \frac{e^{dk-k(z_i-h_0)}\sinh (dk)(2\epsilon_2k \cosh (dk)+2\epsilon_1k \sinh (dk))}{2\epsilon_2\epsilon_1} + \mathcal{O}(\kappa), \\
        A_{II}^{(1,s)}=\frac{(\epsilon_2-\epsilon_1)\pi Q_i(e^{-d-k(z_i-h_0)}(-2(-1+e^{2dk})^2\epsilon_1^2k+\epsilon_2^2(2(-1+e^{4dk})k)+4\epsilon_2\epsilon_1(k+e^{2dk}k))}{4\epsilon_2\epsilon_1k(2\epsilon_2\epsilon_1 \cosh (2dk)+(\epsilon_2^2+\epsilon_1^2) \sinh (2dk))} + \mathcal{O}(\kappa), \\
        B_{II}^{(1,s)}=-\frac{(\epsilon_2-\epsilon_1)\pi Q_ie^{-kd-k(z_i-h_0)}(-4e^{2dk}(\epsilon_2-\epsilon_1)\epsilon_1k-2ke^{4dk}(\epsilon_2+\epsilon_1)^2+2k(\epsilon_2^2-\epsilon_1^2))}{4\epsilon_2\epsilon_1k(2\epsilon_2\epsilon_1 \cosh (2dk)+(\epsilon_2^2+\epsilon_1^2) \sinh (2dk))} + \mathcal{O}(\kappa), \\
        A_{III}^{(1,s)}=-\frac{(\epsilon_2-\epsilon_1)\pi Q_i(e^{-dk-k(z_i-h_0)}(4\epsilon_2\epsilon_1^2(-1+e^{2dk})^2k+4\epsilon_2^2\epsilon_1(-1+e^{4dk})k)}{4\epsilon_2\epsilon_1^2k(2\epsilon_2\epsilon_1 \cosh (2dk)+(\epsilon_2^2+\epsilon_1^2) \sinh (2dk))}+\mathcal{O}(\kappa), \\
        B_{III}^{(1,s)}=0, \\
        A_{IV}^{(1,s)}=-\frac{(\epsilon_2-\epsilon_1)\pi Q_i(e^{-dk-k(z_i-h_0)}(4\epsilon_2\epsilon_1^2(-1+e^{2dk})^2k+4\epsilon_2^2\epsilon_1(-1+e^{4dk})k)}{4\epsilon_2\epsilon_1^2k(2\epsilon_2\epsilon_1 \cosh (2dk)+(\epsilon_2^2+\epsilon_1^2) \sinh (2dk))}+\mathcal{O}(\kappa).\\
    \end{multlined}
\end{AEquation}

With the exception of those applying to the intermembrane region (Region II), these terms do not depend on particle position and will therefore not contribute to the force, adding only a constant to the total energy of the system. Additionally, since the total constant energy term will be equal to $\frac{H}{L_xL_y}\sum_jg_{\vec{m}}^{(1)}(z,zi)Q_j$, this constant will sum to zero for neutral systems. Due to the positional independence of the $\kappa=0$ energetic expressions and the cancellation of the constant term in this case, the corresponding first order sinusoidal $A$ and $B$ terms can be safely ignored.
\subsection{Langevin Dynamics Implementation}
\label{subsection:implementation details}
To implement our solution, we divide the electrostatic potential into a polarization term and a standard electrostatic term as referenced in Section \ref{subsection:general case solutions}:
\begin{AEquation}
    \label{eqn:potential split}
    \phi_{total}=\phi_{s}+\phi_{pol},
\end{AEquation}
where $\phi_{s}$ represents the standard Coulomb contribution to the potential and $\phi_{pol}$ represents the polarization contribution. The standard contribution can be treated by Ewald summation or similar methods.\supercites{santos_simulations_2016}{yeh_ewald_1999} The polarization energy of the system is then defined in the following way:
\begin{AEquation}
    \label{eqn:system energy}
    U_{pol}=\frac{1}{2}\sum_{i}\sum_{j}\phi_{ij},
\end{AEquation}
where $\phi_{ij}$ represents the polarization force felt by particle $j$ due to the presence of particle $i$. From this expression, we determine the total polarization energy of the system using the following equation:

\begin{AEquation}
\begin{multlined}
    \label{eqn:system energy with f functions}
    U_{pol}=\frac{1}{2} \sum_{\vec{m}'} \left[ A_{IV}^{\prime(0)}(f_{01}f_{11}+f_{02}f_{12}+f_{03}f_{13}+f_{04}f_{14})+B_{I}^{\prime(0)}(f_{01}f_{13}+f_{02}f_{14}+f_{03}f_{11}+f_{04}f_{12})\right] \\ +\frac{H}{2}\sum_{\vec{m}''}\left[A_{IV}^{\prime(1,s)}(f_{011}f_{17}+f_{012}f_{18}-f_{09}f_{15}-f_{010}f_{16})+B_{I}^{\prime(1,s)}(f_{011}f_{15}+f_{012}f_{16}-f_{09}f_{17}-f_{010}f_{18}) \right]\\ -\frac{H}{2} \sum_{\vec{m}'}\left[A_{IV}^{\prime(1,c)}(f_{07}f_{13}+f_{08}f_{14}-f_{05}f_{11}-f_{06}f_{12})\right],\\
\end{multlined}
\end{AEquation}

where the primed $A$ and $B$ constants are identical to the $A$ and $B$ constants defined in Equations \ref{eqn:values of zero order constants}, \ref{eqn:values of first order sinusoidal constants}, and \ref{eqn:values of first order constant constants} with the dependencies on $z_i$ and $Q_i$ removed. The primed summation over $\vec{m}$ indicates exclusion of the $\vec{m}=0$ term in the summation; the double primed summation indicates exclusion of the $\vec{m}=0$ and $\vec{m}+\vec{q}=0$ terms. The $f$ functions are defined as follows:
\begin{subequations}
\begin{AEquation}
    \begin{multlined}
        \label{eqn:f function definitions part 1}
        \\f_{01}=\sum_{i\in l}Q_ie^{-2kh_0+kz_i}\cos \left(\frac{2\pi m_x}{L_x}x_i+\frac{2\pi m_y}{L_y}y_i\right), \\
        f_{02}=\sum_{i\in l}Q_ie^{-2kh_0+kz_i}\sin \left(\frac{2\pi m_x}{L_x}x_i+\frac{2\pi m_y}{L_y}y_i\right), \\
        f_{03}=\sum_{i\in r}Q_ie^{2kh_0-kz_i}\cos \left(\frac{2\pi m_x}{L_x}x_i+\frac{2\pi m_y}{L_y}y_i\right), \\
        f_{04}=\sum_{i\in r}Q_ie^{2kh_0-kz_i}\sin \left(\frac{2\pi m_x}{L_x}x_i+\frac{2\pi m_y}{L_y}y_i\right), \\
        f_{05}=\sum_{i\in l}Q_ie^{-2kh_0+k z_i}\cos \left(\frac{2\pi m_x}{L_x}x_i+\frac{2\pi m_y}{L_y}y_i\right) \cos (q_x x_i + q_y y_i), \\
        f_{06}=\sum_{i\in l}Q_ie^{-2kh_0+k z_i}\sin\left(\frac{2\pi m_x}{L_x}x_i+\frac{2\pi m_y}{L_y}y_i\right)\cos(q_x x_i + q_y y_i), \\
        f_{07}=\sum_{i\in r}Q_ie^{2kh_0-k z_i}\cos\left(\frac{2\pi m_x}{L_x}x_i+\frac{2\pi m_y}{L_y}y_i\right)\cos(q_x x_i + q_y y_i), \\
        f_{08}=\sum_{i\in r}Q_ie^{2kh_0-k z_i}\sin\left(\frac{2\pi m_x}{L_x}x_i+\frac{2\pi m_y}{L_y}y_i\right)\cos(q_x x_i + q_y y_i),\\
        f_{09}=\sum_{i\in l}Q_ie^{-(k+\kappa)h_0+kz_i}\cos\left(\frac{2\pi m_x}{L_x}x_i+\frac{2\pi m_y}{L_y}y_i\right), \\
        f_{010}=\sum_{i\in l}Q_ie^{-(k+\kappa)h_0+kz_i}\sin\left(\frac{2\pi m_x}{L_x}x_i+\frac{2\pi m_y}{L_y}y_i\right), \\
        f_{011}=\sum_{i\in r}Q_ie^{(k+\kappa)h_0-kz_i}\cos\left(\frac{2\pi m_x}{L_x}x_i+\frac{2\pi m_y}{L_y}y_i\right), \\
        f_{012}=\sum_{i\in r}Q_ie^{(k+\kappa)h_0-kz_i}\sin\left(\frac{2\pi m_x}{L_x}x_i+\frac{2\pi m_y}{L_y}y_i\right), \\
    \end{multlined}
\end{AEquation}
and
\begin{AEquation}
    \begin{multlined}
        \label{eqn:f function definitions part 2}
        \\
        f_{11}=\sum_{i\in l}Q_ie^{kz_i}\cos\left(\frac{2\pi m_x}{L_x}x_i+\frac{2\pi m_y}{L_y}y_i\right), \\
        f_{12}=\sum_{i\in l}Q_ie^{kz_i}\sin\left(\frac{2\pi m_x}{L_x}x_i+\frac{2\pi m_y}{L_y}y_i\right), \\
        f_{13}=\sum_{i\in r}Q_ie^{-kz_i}\cos\left(\frac{2\pi m_x}{L_x}x_i+\frac{2\pi m_y}{L_y}y_i\right), \\
        f_{14}=\sum_{i\in r}Q_ie^{-kz_i}\sin\left(\frac{2\pi m_x}{L_x}x_i+\frac{2\pi m_y}{L_y}y_i\right), \\
        f_{15}=\sum_{i\in l}Q_ie^{\kappa z_i}\cos\left(\frac{2\pi m_x}{L_x}x_i+\frac{2\pi m_y}{L_y}y_i\right)\cos(q_x x_i + q_y y_i), \\
        f_{16}=\sum_{i\in l}Q_ie^{\kappa z_i}\sin\left(\frac{2\pi m_x}{L_x}x_i+\frac{2\pi m_y}{L_y}y_i\right)\cos(q_x x_i + q_y y_i), \\
        f_{17}=\sum_{i\in r}Q_ie^{-\kappa z_i}\cos\left(\frac{2\pi m_x}{L_x}x_i+\frac{2\pi m_y}{L_y}y_i\right)\cos(q_x x_i + q_y y_i), \\
        f_{18}=\sum_{i\in r}Q_ie^{-\kappa z_i}\sin\left(\frac{2\pi m_x}{L_x}x_i+\frac{2\pi m_y}{L_y}y_i\right)\cos(q_x x_i + q_y y_i),\\
    \end{multlined}
\end{AEquation}
\end{subequations}
where $i\in l$ and $i \in r$ represent summations over particles on the left and right sides of the membrane respectively. Additionally, we note that all complex exponential expressions are replaced with their real components. Implementation of Equation \ref{eqn:system energy with f functions} leads to extremely efficient computations of the polarization force; the complexity of such calculations is of $\mathcal{O}(N)$ where $N$ is the number of particles in the system.\par
\printbibliography
\pagebreak
\section{Supplementary Material}
\subsection{Self Energy vs Lateral Position}
\begin{figure*}[h]
    \centering
    \includegraphics[width=\textwidth]{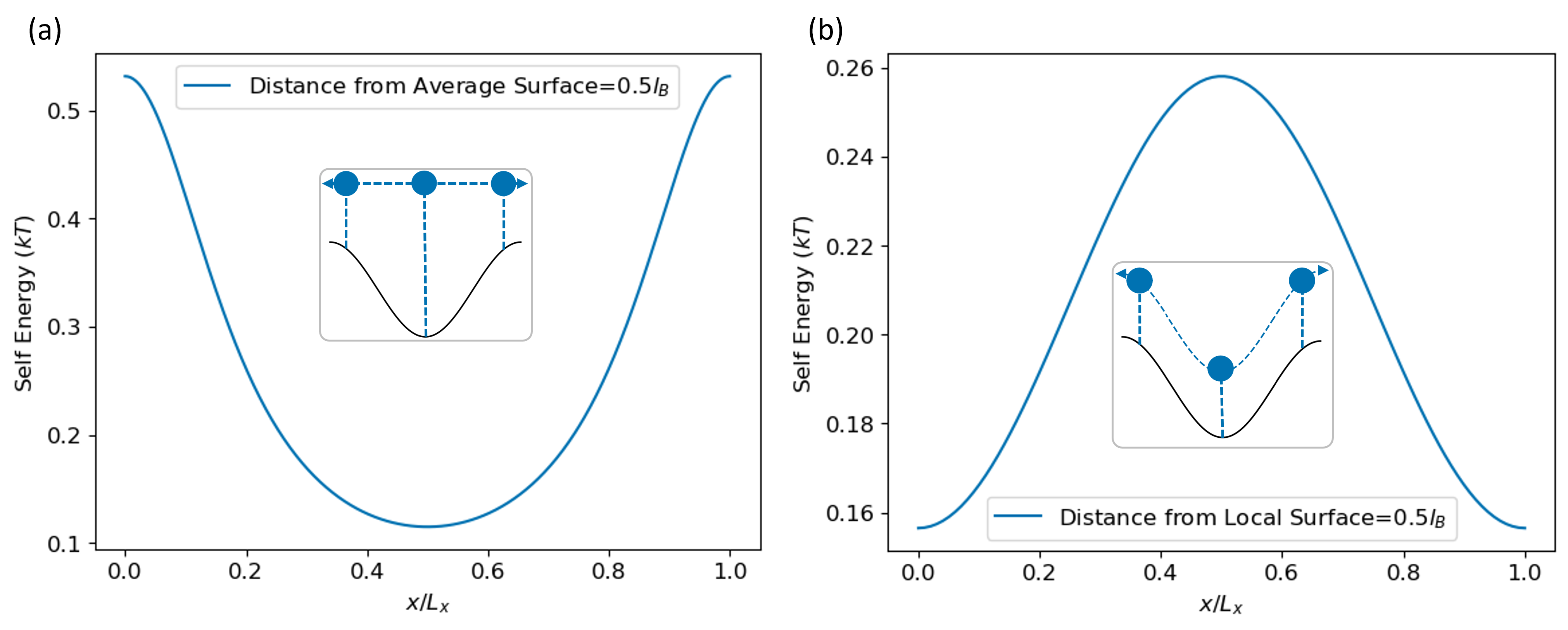}
    \caption{(a)  Self-energy of a monovalent charged particle vs its $x$ position for a constant $z$ position above a sinusoidally deformed interface. The self-energy of the particle follows the contour of the membrane, increasing above peaks and decreasing above troughs due to changes in particle proximity to induced membrane charge. (b) Self-energy of a monovalent charged particle vs its $x$ position while maintaining a constant height above a sinusoidally deformed surface. The self-energy of the particle is an inverse of the membrane contour due to curvature effects dominating when the local distance to the surface is held constant. Note that this particular membrane is structured such that there are peaks at either end and a trough in the middle. The insets show the particle location relative to the curved membrane surface as it is translated in the $x$ direction. The deformation is described by the parameters $H=0.25l_B$ and $\lambda=5l_B$.}
    \label{fig:self energy vs x combined}
\end{figure*}
\pagebreak
\subsection{\texorpdfstring{Peak and Trough Charge Densities for $\lambda=5l_B$ System}{Peak and Trough Charge Densities for lambda=5 System}}
\begin{figure*}[h]
    \centering
    \includegraphics[width=\textwidth]{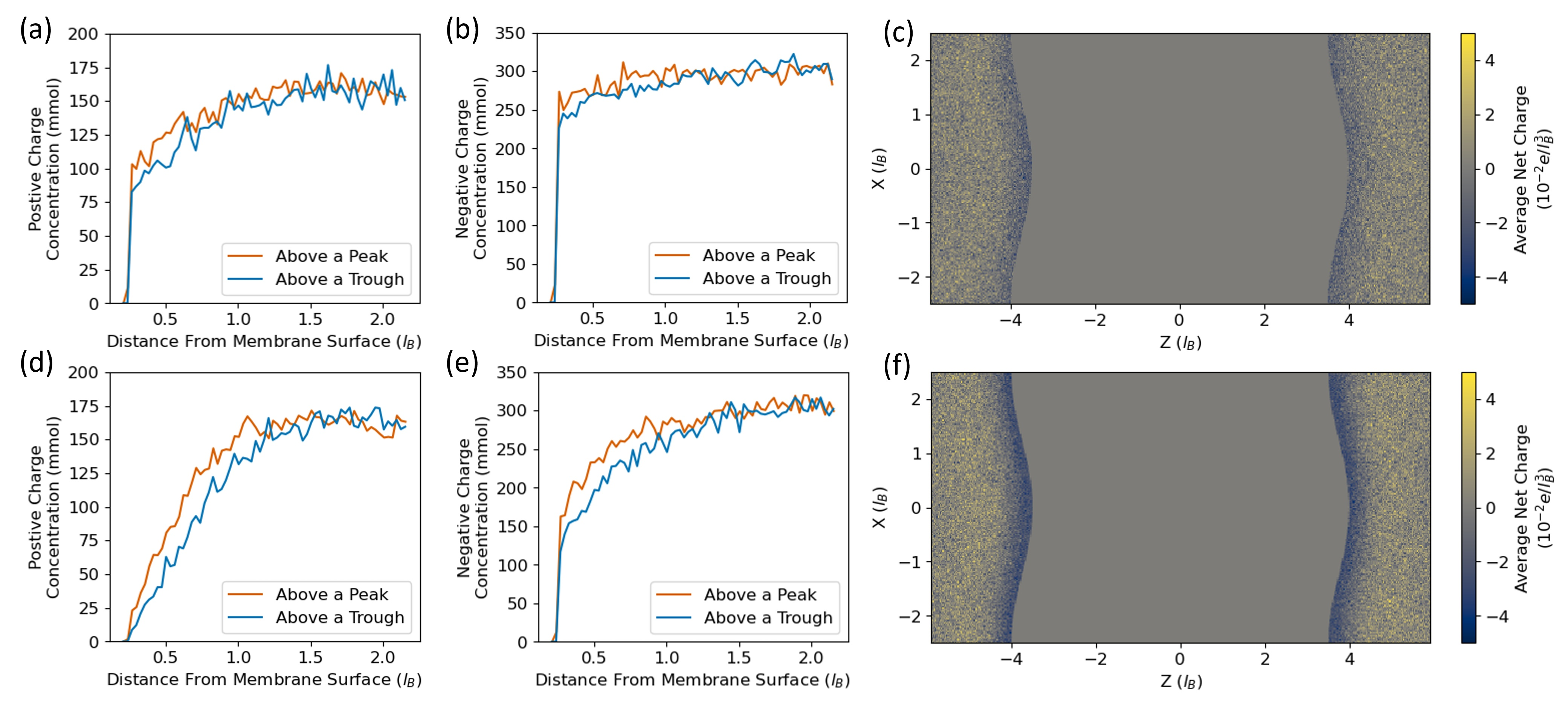}
    \caption{Divalent positive and monovalent negative charge densities above peaks and troughs of a sinusoidally deformed surface with $\lambda=5l_B$. Top: (a) positive and (b) negative charge densities above the surface without dielectric mismatch, comparing densities above peaks (orange) and troughs (blue). (c) Two-dimensional plot of the net charge density near the interface. Bottom: (d) Positive and (e) negative charge  densities above peaks (orange) and troughs (blue) of a surface with dielectric mismatch. (f) Two-dimensional plot of the net charge density near the interface incorporating this dielectric mismatch. Dielectric mismatch tends to lead to a net negative charge at the interface due to the stronger repulsion of divalent positive ions than monovalent negative ions.}
    \label{fig:net charge lambda 10}
\end{figure*}
\pagebreak
\subsection{\texorpdfstring{Charge Density Maps for $\lambda=2.5l_B$ System Without Dielectric Mismatch}{Charge Density Maps for lambda=2.5 System Without Dielectric Mismatch}}
\begin{figure*}[h]
    \centering
    \includegraphics[width=\textwidth]{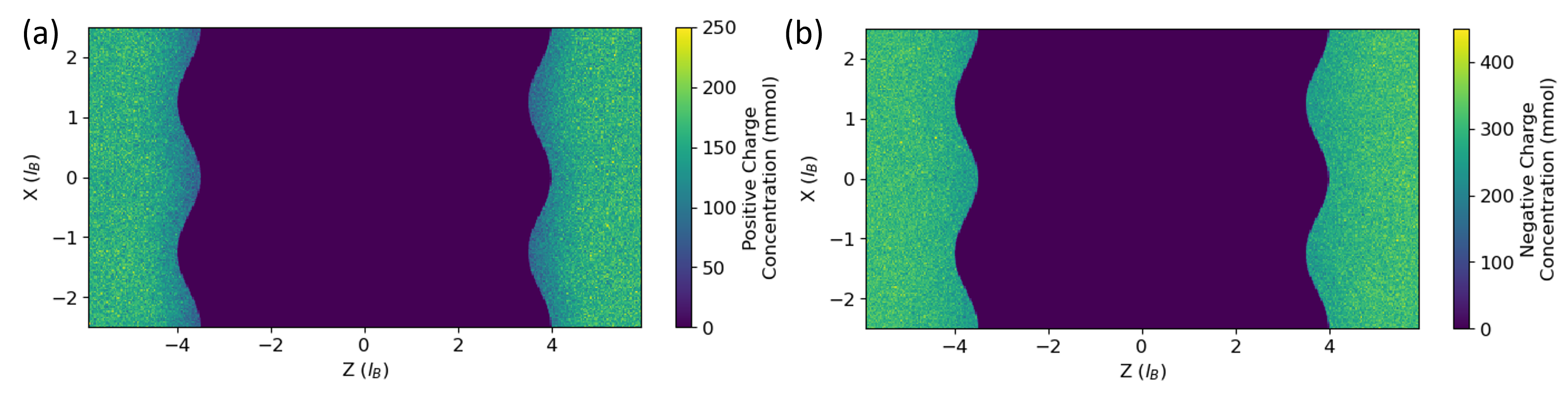}
    \caption{(a) Divalent positive and (b) monovalent negative charge densities near a sinusoidally deformed interface with $\lambda=2.5l_B$, $H=0.25l_B$, and no dielectric mismatch. Some depletion of positive ions from the interface can be seen; this can be explained by an attraction of interfacial ions to the ionic bulk.}
    \label{fig:charge densities lambda 2.5}
\end{figure*}
\pagebreak
\subsection{\texorpdfstring{Simulation Results for Alternative $\epsilon_2$}{Simulation Results For Alternative epsilon2}}
While dielectric mismatch is important to membranes and membrane-like systems, our methods are equally interesting to other dielectric systems such as clays. In Figure \ref{fig:clay results} below, we present simulation results for dielectric mismatch typical of a water-clay interface. We simulate a system with a deformation of amplitude $H=0.25 l_B$ and wavelength $\lambda=2.5l_B$. All simulation parameters save the dielectric constants are identical to the other simulations in this work. We set $\epsilon_1=80$ and $\epsilon_2=20$ to mimic a typical water-clay interface.\supercites{hubbard_estimation_1997}{zhang_frequency_2020} 
\begin{figure*}[h]
    \centering
    \includegraphics[width=\textwidth]{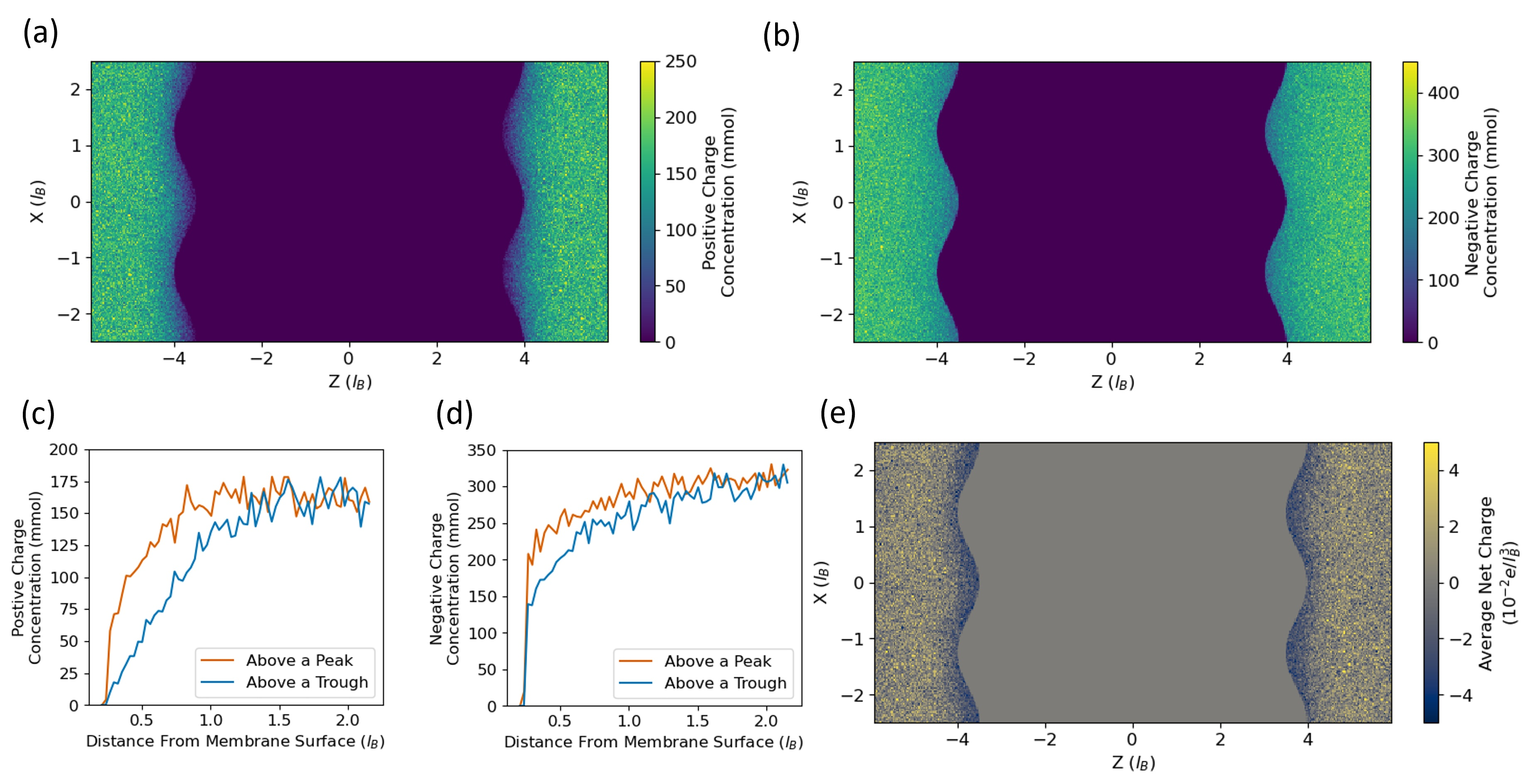}
    \caption{Top: (a) Divalent positive and (b) monovalent negative charge densities near a sinusoidally deformed dielectric interface with amplitude $H=0.25 l_B$ and wavelength $\lambda=2.5l_B$. Dielectric mismatch mimics a typical water-clay interface with $\epsilon_1=80$ and $\epsilon_2=20$. Under these conditions, the effect of the dielectric mismatch is still noticeable and results in decreased charge density near the interface; however the effect is less obvious than in the membrane systems with $\epsilon_2=2$. This is likely due to the reduced dielectric mismatch. Bottom: Charge densities of (c) divalent positive and (d) monovalent negative ions above peaks and troughs of the surface. (e) Net charge density near the water-clay interface. The effects of the curved surface are reduced from the membrane case, but still noticeable with a notable relative ionic preference for peaks over troughs.}
    \label{fig:clay results}
\end{figure*}
\pagebreak
\subsection{Numerical Validation}
\begin{figure*}[h]
    \centering
    \includegraphics[width=\textwidth]{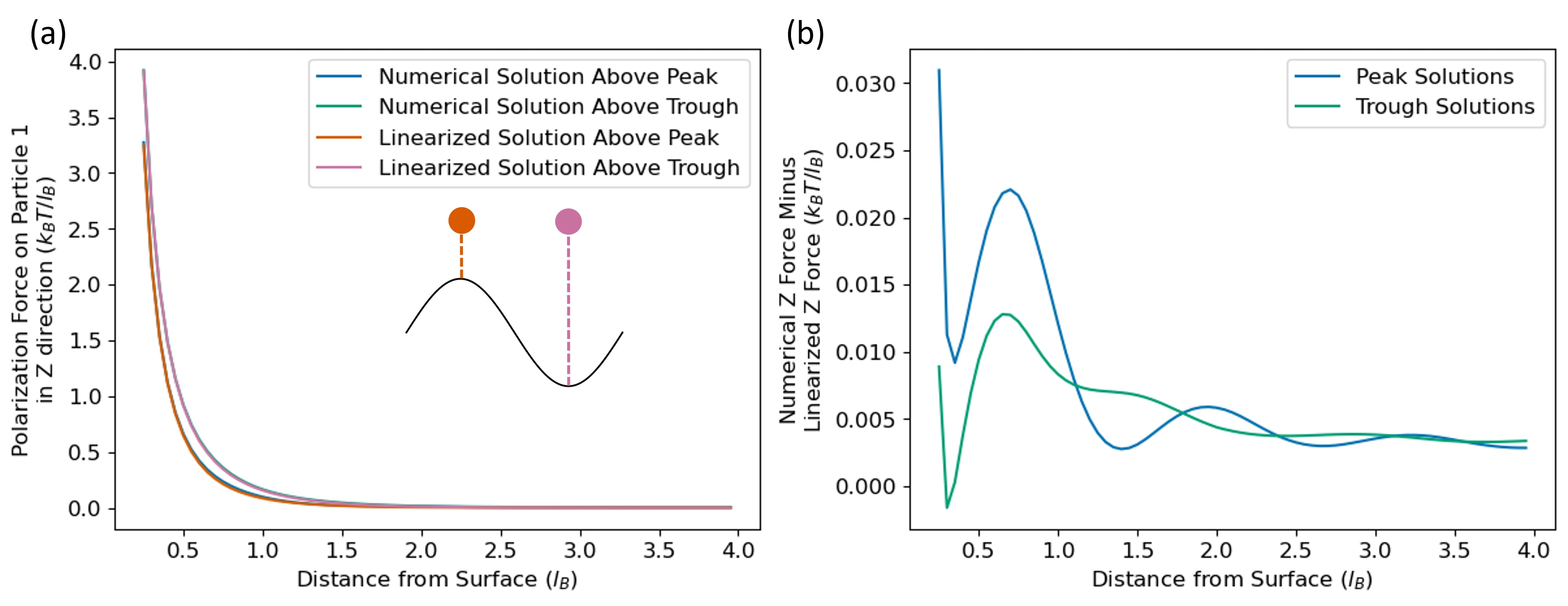}
    \caption{(a) Plot of the z-component of force on a test particle vs its distance from the membrane surface. Results are shown for particles above peaks and above troughs, and are furthermore shown for both the linearized method developed in this current work and a numerical method which computes the induced charge at the dielectric interface by treating the surface as a triangular mesh. The numerical and linear solutions overlap to the point of being largely indistinguishable; this suggests that our method is valid even for particles close to the surface. (b) Difference between the linearized and numerical forces on a test particle vs its distance from the membrane surface. Differences are small throughout the range of values sampled, growing largest near the surface. This is reasonable and expected as both our model and our chosen numerical method have the most difficulty describing interactions very close to the surface. Despite some differences near the surface, the models appear to agree extremely well with one another, emphasizing the validity of our linear approximation approach.}
    \label{fig:numerical validation}
\end{figure*}

We validate our numerical method by comparing the forces in the z direction on a test particle near a sinusoidally deformed interface of amplitude $H=0.5 l_B$ and $\lambda=2.5 l_B$. For these simulations, all parameters are identical to the rest of the simulations performed in this work save for the number of particles which is set to 2. The system contains one cation with charge $+1$ and one anion with charge $-1$. The cation is positioned above either a peak or a trough, while the anion is fixed at a set distance from the cation. We calculate polarization force profiles for the cation by moving it in the $+z$ direction over a range of positions, starting very close to the membrane surface and subsequently moving further away. We determine the polarization contribution to the force by taking the difference of the force on the test cation with and without dielectric mismatch. Our numerical method works by numerically integrating the charge on the surface using the Poisson equation; the surface is represented as a 11730-point triangular mesh.\supercites{nguyen_incorporating_2019}{foster_gpu_2020}{nguyen_gpu-accelerated_2017}{nguyen_accelerating_2015}{brown_implementing_2013}{brown_implementing_2012}{brown_implementing_2011} The results of this comparison are shown in Figure \ref{fig:numerical validation}.\par
It is worth additionally noting that our method does have limits on the amplitudes of the deformations which can be accommodated. Based on the boundary conditions of the unperturbed solution, the results fail to converge for particles which sit inside the inner region of the system (Region II as defined in Figure 1 of the main paper). The solution here is qualitatively different and as a result the approximation breaks down. As a result, our approximation in its current form can handle deformation amplitudes of up to half the particle diameter. While this is certainly somewhat limiting, it is very useful for describing larger ions such as those comprising ionic liquids; moreover, the results from these simulations can provide intuition for the design of other, more drastically deformed systems.
\end{document}